\documentclass[pdflatex,sn-standardnature]{sn-jnl}
\usepackage{graphicx}
\usepackage{amsmath,amssymb,amsfonts}%
\usepackage{amsthm}%
\usepackage{nicefrac,xfrac}
\usepackage{siunitx}
\usepackage[utf8]{inputenc} 
\usepackage[T1]{fontenc}    
\usepackage{lmodern}        
\usepackage[T1]{fontenc}
\usepackage{textcomp}
\hypersetup{bookmarksnumbered=true}
\usepackage{caption} 
\usepackage{chngcntr} 

\newcommand{\SUPFIG}[1] {Supplementary Figure~\ref{#1}}

\graphicspath{{fig/2nd-round/}}

\definecolor{darkblue}{cmyk}{1, 1, 0, 0}
\hypersetup{colorlinks=true,urlcolor=darkblue,citecolor=darkblue,linkcolor=darkblue}

\begin{document}

\title{Co-precession of a curved jet and compact accretion disk in M87}

\author*[1,2]{\fnm{Yuzhu} \sur{Cui}}
\email{\textcolor{black}{yuzhu\_cui77@163.com}}
\equalcont{Both authors contributed equally to this work. The authors are listed in alphabetical order by their family names.}
\affil*[1]{Institute of Astrophysics, Central China Normal University, Wuhan 430079, China}
\affil[2]{Research Center for Astronomical Computing, Zhejiang Lab, Hangzhou 311100, China}

\author*[3]{\fnm{Weikang} \sur{Lin}}
\email{\textcolor{black}{weikanglin@ynu.edu.cn}}
\equalcont{Both authors contributed equally to this work. The authors are listed in alphabetical order by their family names.}
\affil*[3]{South-Western Institute For Astronomy Research, Yunnan University, Kunming 650500, Yunnan, P. R. China}

\abstract{
Observational constraints on the configuration of the black hole (BH)-accretion disk-jet system are crucial to understanding BH spin, accretion disk physics, and jet formation. The recently reported variation in the M87 jet position angle (PA) provides a novel avenue to explore these long-standing issues. The observed $\sim$11-year periodicity, spanning over two cycles, is consistent with the Lense–Thirring (LT) precession of a compact, tilted accretion disk. However, how such a compact region decouples from the larger-scale accretion flow remains an open question in current numerical simulations. The jet precession challenges the traditional view of a strictly collimated jet, revealing a subtle curvature in the jet's inner regions that dynamically links the jet to the spinning BH and successfully accounts for its unexpectedly wide inner projected profile. While continued long-term observations are needed to distinguish coherent precession from stochastic fluctuations in the disk or jet orientation, these results open a new window for probing BH systems through coordinated multi-scale observations and follow-on theoretical models.}


\maketitle

\section*{Main}
The nearby radio galaxy M87 is the first observational record of an astrophysical jet reported in 1918~\cite{curtis1918}. Despite extensive studies, several fundamental aspects of its central engine remain uncertain. While the SMBH at the core of M87 is known to accrete at a low rate through a geometrically thick, radiatively inefficient accretion flow~\cite{igumenshchev2003, yuan2014, eht52019, eht82021}, the exact structure, size, and orientation of the accretion disk are poorly constrained by current observations. Additionally, determining the SMBH spin in M87 has proven challenging. Traditional techniques like X-ray reflection spectroscopy~\cite{reynolds2013, Risaliti2013} and thermal continuum fitting~\cite{McClintock2006, McClintock2014} are limited by the thick geometry and low accretion rate of the disk, making them unsuitable for M87. The Event Horizon Telescope (EHT) revolutionized our understanding of M87’s SMBH by capturing the first direct image of photon orbits at 230 GHz. This breakthrough provided robust constraints on the SMBH’s mass~\cite{eht12019, eht62019}, offering unprecedented insights into the immediate vicinity of the event horizon. Beyond the mass determination, the EHT findings also excluded models with zero spins for the SMBH, as they failed to account for the observed jet power~\cite{eht52019}. As relativistic jets in active galactic nuclei (AGNs) are intrinsically linked to the SMBH and the surrounding accretion disk~\cite{blandford1977, blandford1982}, probing the inner regions of the BH-disk-jet system is essential for further understanding physical conditions and relativistic effects at play near SMBHs. M87, renowned for its powerful AGN with an extended relativistic jet, provides an ideal laboratory for this topic (\cite{hada2024}, and references therein).

Recent high-resolution observations of the M87 jet at milliarcsecond (mas) and sub-mas scales have provided crucial new insights into the above challenges~\cite{Lu:2023bbn, cui2023}. Notably, Cui et al. (2023) identified a periodic variation in the jet position angle (PA), with a cycle of approximately 11 years, interpreted as Lense-Thirring (LT) precession due to a tilted accretion disk~\cite{cui2023}. LT precession, a relativistic effect caused by frame-dragging near a spinning SMBH~\cite{lense1918}, offers a unique opportunity to infer properties of the BH-disk-jet system that are otherwise difficult to measure directly. In this work, we demonstrate that the observed jet precession provides stringent constraints on the SMBH spin and accretion disk parameters. In addition, the precession implies that the jet is not perfectly collimated, particularly in regions close to the central SMBH, addressing the projected jet width discrepancy between the observations and simulations reported by~\cite{Lu:2023bbn} without unexpected emission component outside the Blandford–Znajek (BZ) mechanism~\cite{blandford1977}. 

\section{The SMBH spin and accretion disk size}
\begin{figure}
    \centering
    \includegraphics[width=0.8\linewidth]{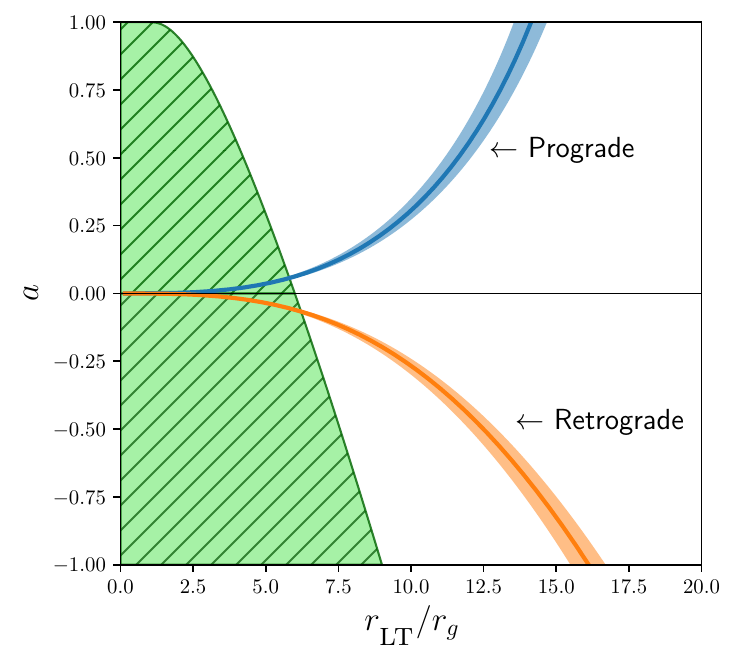}
    \caption{\textbf{Constraints on the M87 BH spin parameter ($a$) and effective disk LT radius ($r_{\rm LT}$).} The constraint is set by the observed jet precession period and SMBH mass. The green shaded region is excluded based on the conservative requirement that $r_{\textsc{lt}}>r_{\textsc{isco}}$. The blue (orange) curve corresponds to a prograde (retrograde) disk.}
    \label{fig:period-a-rlt}
\end{figure}

Due to computational cost and technical issues, comprehensive numerical studies of tilted and geometrically thick accretion disks were not possible until recently~\cite{fragile2005, fragile2007}. In particular, the main body of the disk undergoes LT precession~\cite{fragile2005, fragile2007}, and the jet follows the disk and precesses with the same period~\cite{mcKinney2013, cui2023}. The precession period can be calculated by
\begin{equation}\label{eq:LT-precession-effective-higher-order}
    T_{\rm prec}\simeq\frac{\pi c^3 r_{\textsc{lt}}^3}{aG^2M^2}\left[1-\frac{3}{4}a(r_{\rm g}/r_{\textsc{lt}})^{1/2}\right]^{-1}\,,
\end{equation}
where we have taken into account the sense of the accretion flow orbit concerning the central SMBH spin and the higher-order corrections~\cite{kato1990, Franchini:2015wna} (see Methods). In the above, $a\equiv Jc/GM^2\in[-1,1]$ is the dimensionless spin parameter and a positive (negative) $a$ corresponds to a prograde (retrograde) disk, $M$ is the SMBH mass, and $r_{\rm g}=GM/c^2$ is the gravitational radius, $J$ is the spin angular momentum, $G$ is the gravitational constant, $c$ is the speed of light, and $r_{\textsc{lt}}$ is an effective radius that depends on the mass distribution of the part of the disk that exhibits a coherent precession (denoted as the ``precessing disk''). For M87, the reported precession period $T_{\rm prec}=11.24\pm0.47$ years~\cite{cui2023} and the SMBH mass $M_{\rm M87}=(6.5\pm0.7)\times 10^9\,M_\odot$ (where $M_\odot$ is the solar mass)~\cite{eht12019} put a constraint on the $a$--$r_{\textsc{lt}}$ space which is shown in Fig.~\ref{fig:period-a-rlt}. 

The relatively short precession period puts a tight constraint on the size of the precessing disk and requires $r_{\textsc{lt}}\leq14.1\pm 0.6\,r_{\rm g}$ for the prograde case and $r_{\textsc{lt}}\leq16.1\pm 0.6\,r_{\rm g}$ for the retrograde case. The equal sign corresponds to the case of maximum rotation, i.e. $a=1$ or $a=-1$. For the same $\lvert a\rvert$, the required $r_{\textsc{lt}}$ is somewhat smaller for a prograde disk compared to a retrograde disk. A smaller $\lvert a\rvert$ requires a smaller $r_{\textsc{lt}}$ for both the prograde and retrograde cases. Thus, the precessing disk is constrained to be rather compact in the sense that the effective LT radius is at most around $15\,r_{\rm g}$.

The constraint on $a$ is highly sensitive to $r_{\textsc{lt}}$. Assuming $r_{\textsc{lt}}>r_{\textsc{isco}}$ where $r_{\textsc{isco}}$ is the innermost stable circular orbit~\cite{bardeen1972}, we establish a lower limit of $a\gtrsim0.06$~\cite{cui2023}. This result is consistent with that given in~\cite{Wei:2024cti} despite the difference in the jet-disk co-precessing scenario (see Methods). Although it provides a weak constraint on the value of $a$, the non-spinning scenario is successfully excluded, consistent with the findings of~\cite{eht52019}. 

\begin{figure}
    \centering
    \includegraphics[width=\linewidth]{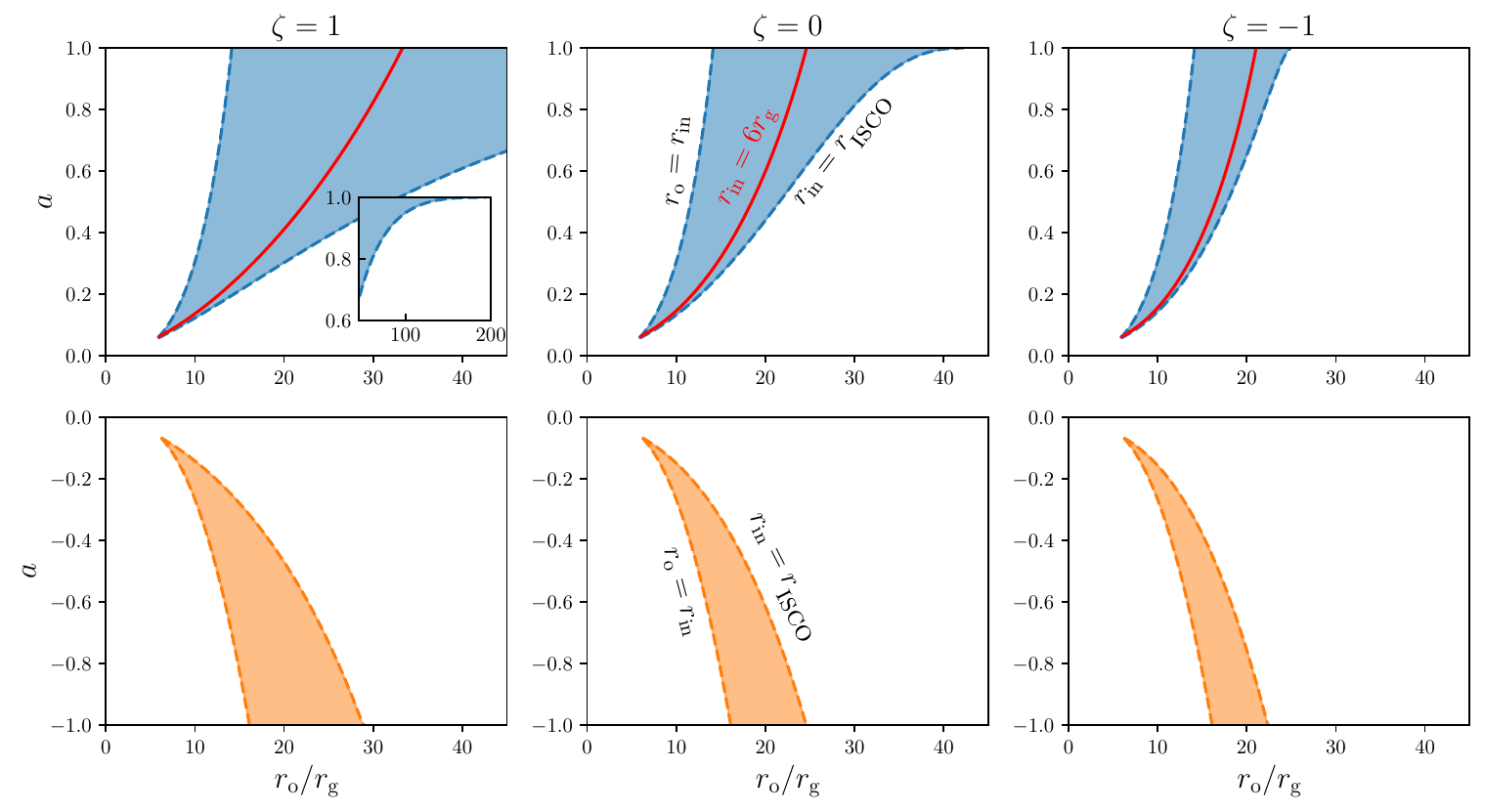}
    \caption{\textbf{Constraints on the M87 BH spin and accretion disk morphology in the $a$--$r_{\rm o}$ (outer radius of the precessing disk) parameter space.} The upper (lower) panels are for prograde (retrograde) disks. A larger profile index $\zeta$ represents a disk that is more concentrated inwards. The shaded areas enclose the allowed region where $r_{\textsc{isco}}<r_{\rm in}<r_{\rm o}$. The red curves represent the cases where $r_{\rm in}=6\,r_{\rm g}$ motivated by the finding in simulations that the inner radius for prograde tilted disks is insensitive to the black hole spin~\cite{fragile2009} for weakly magnetized disk, serving as a guiding and limiting case.}
    \label{fig:constraint-details}
\end{figure}

Considering the extended size of the precessing disk, the assumption $r_{\textsc{lt}}>r_{\textsc{isco}}$ is highly conservative. To analyze the structure of the precessing disk, we follow~\cite{Liu:2002wu, fragile2007} and adopt a power-law surface density profile with $\sigma(r)\propto r^{-\zeta}$. A positive (negative) $\zeta$ represents a disk that is more concentrative inwards (outwards), and $\zeta=0$ represents a constant surface density. 
The inner and outer radii of the precessing disk are denoted by $r_{\rm in}$ and $r_{\rm o}$, respectively. These two characteristic radii provide an effective description of the precessing disk’s overall extent. With these settings, the effective LT radius is expressed as (\cite{fragile2007}, see Methods),
\begin{equation}\label{eq:LT-radius-fragile}
    r_{\textsc{lt}} = r_{\rm o}^{\frac{5-2\zeta}{6}}r_{\rm in}^{\frac{1+2\zeta}{6}}\left[\frac{1+2\zeta}{5-2\zeta}\frac{1-(r_{\rm in}/r_{\rm o})^{\nicefrac{5}{2}-\zeta}}{1-(r_{\rm in}/r_{\rm o})^{\nicefrac{1}{2}+\zeta}} \right]^{\nicefrac{1}{3}}\,.
\end{equation}
It is justified to assume $r_{\rm in}>r_{\textsc{isco}}$, which sets a lower limit of $a$ for a given $r_{\rm o}$ and profile index $\zeta$. In addition, $r_{\rm in}<r_{\rm o}$ is geometrically required, which sets an upper limit of $a$ for a given $r_{\rm o}$. Taking $T_{\rm prec}=11.24$\,years~\cite{cui2023} and $M_{\rm M87}=6.5\times10^{9}\,M_\odot$~\cite{eht12019}, the resultant constraints on the $a$--$r_{\rm o}$ space, capturing $r_{\textsc{isco}}<r_{\rm in}<r_{\rm o}$, are shown by the shaded areas in Fig.~\ref{fig:constraint-details}. These constraints are presented for three selected values of $\zeta$, with different senses of the disk orbit considered separately. 

The general relativistic magnetohydrodynamic (GRMHD) simulations indicate that the disk surface density is nearly constant ($\zeta=0$)~\cite{fragile2007}, which we denote as the fiducial case. For this scenario, the maximal outer radius is $\sim42\,r_{\rm g}$ ($25\,r_{\rm g}$) for a prograde (retrograde) disk, further suggesting a rather compact precessing disk. The constraint of its size is sensitive to the profile index $\zeta$. In general, a larger $\zeta$ allows a larger outer radius because a mass profile more concentrated inwards permits a larger $r_{\rm o}$ to achieve the same $r_{\textsc{lt}}$. For a fairly inward-concentrated and prograde disk with $\zeta=1$, the outer radius can be up to $\sim194\,r_{\rm g}$ for a maximally spinning SMBH.  

For aligned accretion disks, prograde cases generally lead to a larger precessing disk compared to retrograded cases because $r_{\textsc{isco}}$ is smaller, allowing $r_{\rm in}$ to be smaller. For tilted accretion disks, this may not be true. Simulations show that $r_{\rm in}$ is nearly independent of the spin for a modest tilt in a prograde case~\cite{fragile2009} where a weak magnetic field was adopted. This suggests that even for a highly spinning black hole, $r_{\rm in}$ might not shrink with $r_{\textsc{isco}}$, and the tight constraint on the outer radius might not be relaxed. Although similar studies for cases with moderate or strong magnetic fields are not yet available, we consider this scenario as both a guiding and limiting case. We illustrate the cases with $r_{\rm in}=6\,r_{\rm g}$ by the red curves in the upper panels of Fig.~\ref{fig:constraint-details}. Even though retrograde disk cases have not been studied in simulations, it is reasonable to assume that $r_{\rm{in}} > r_{\textsc{isco}}$ still holds. Therefore, the conclusion that $r_{\rm{in}}$ is nearly constant might not apply to retrograde disks.

A spinning SMBH induces an asymmetric shape in the BH ring~\cite{eht52019,sheperd2023} and creates a distinctive rotational pattern in the polarization map~\cite{Palumbo_2020,Chael_2023}. Future EHT observations with next-generation upgrades~\cite{sheperd2023} that resolve these features will be essential for placing stringent constraints on the SMBH’s spin~\cite{Ricarte:2022kft}, breaking the spin-disk size degeneracy described above.

\section{The jet structure and the inner jet width}\label{sec:innermost-jet}
\begin{figure}
    \centering
    \includegraphics[width=0.99\linewidth]{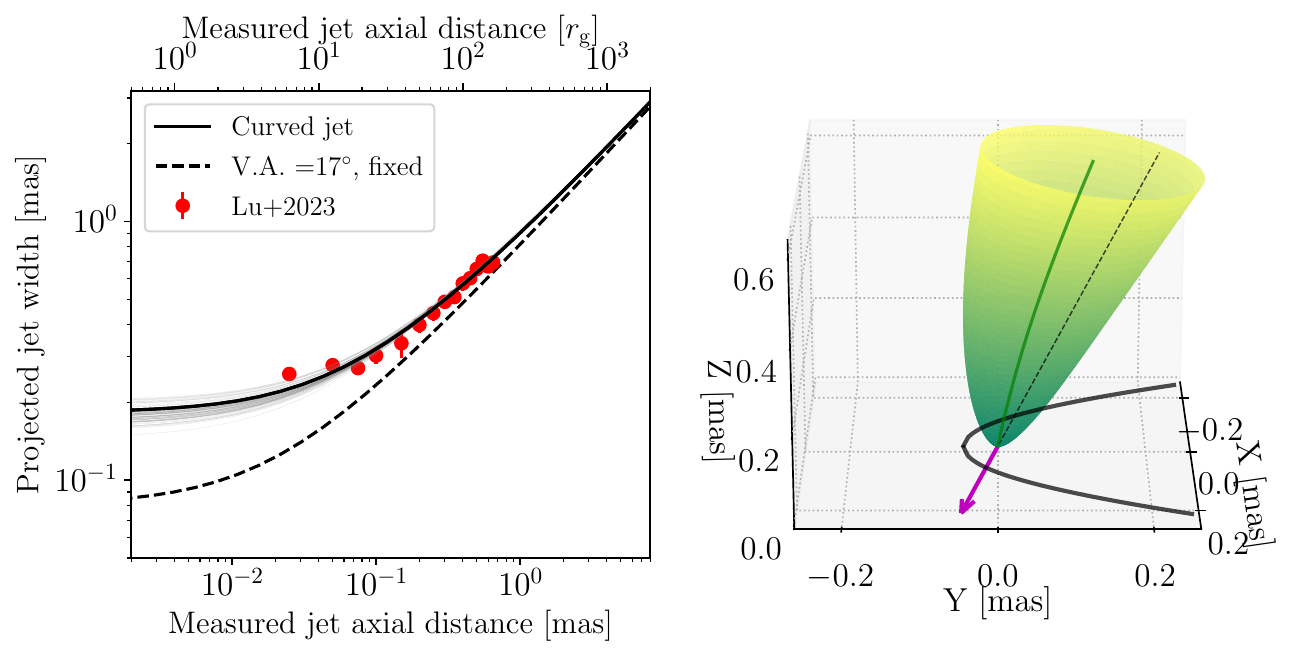}
    \caption{\textbf{Curved jet at the inner region.} {\it Left}: The red points are the projected jet width profile adopted from Lu et al. (2023) which were measured from the observation with GMVA + ALMA + GLT at 86\,GHz in 2018~\cite{Lu:2023bbn}. The dashed curve shows the projected jet width of the collimated jet with $a=0.9$ viewed from a fixed viewing angle $\phi = 17^{\circ}$~\cite{nakamura2018, Lu:2023bbn}, while the solid curve indicates the corresponding projected jet width of a curved jet with the best-fit parameters. The light curves are predictions with randomly selected parameter samples obtained from a Markov Chain Monte Carlo (MCMC) analysis and represent the uncertainty of the fitting. The precession angle at mas scales, $(1.25\pm0.18)^\circ$~\cite{cui2023}, was incorporated into the MCMC analysis. {\it Right}: 3D configuration of the jet structure and the corresponding projected profile on the sky plane. $Z$-axis is set along with LOS. The positive $z$ represents the direction pointing towards the Earth and the $X$--$Y$ plane indicates the sky plane. The green cone is the jet and the green line is the jet axis. The magenta arrow denotes the BH spin, which is aligned with the black dashed line representing the precession axis.}\label{fig:innerjet-fitting}
\end{figure}

The second important implication concerns the jet’s structure and opening angle near BH. The observed periodic PA variations of the M87 jet challenge the traditional view of it being strictly collimated and reveal a jet precession around the central SMBH's spin axis~\cite{cui2023}. The inferred precession angle reflects the angle between the SMBH spin and the tangential direction of the jet axis at the given scales. Complemented with observations showing that the jet's structure is stable in the long-term monitoring and becomes well-collimated at larger distances~\cite{asada2012, nakamura2018, mwl2021, nikonov2023, MWL2024}, the small precession angle observed at the mas scales ($\sim 1.25^\circ$) indicates that the SMBH spin is nearly aligned with the large-scale jet. Simulations of tilted accretion disks show that the inner precessing jets are perpendicular to the coherently precessing disk~\cite{fragile2007, liska2018}. To achieve a consistent description of the above, the jet is expected to curve gradually and its precession angle decreases with the distance from the central SMBH. Hence, we propose the following ansatz for the precession angle $\psi(\rho)$ as a function of the jet’s intrinsic length $\rho$:
\begin{equation}\label{eq:precession-angle-parameterization}
    \psi(\rho) = \psi_{\rm in}\times\left[0.5-\frac{1}{\pi}\arctan(\frac{\log_{10}\rho-\log_{10}\rho_{\rm t}}{\Delta})\right]\,,
\end{equation}
where $\psi_{\rm in}$ is the precession angle at the inner region of the jet, $\rho_{\rm t}$ denotes the transition location, and $\Delta$ represents the transition width. This curved jet configuration with the large-scale jet aligning with the SMBH spin is a plausible outcome. While the inner jet is tightly coupled to the randomly oriented disk at small scales~\cite{mcKinney2013, liska2018}, both the SMBH spin and the large-scale jet are strongly influenced by the large-scale ambient medium. The SMBH spin is determined by the net angular momentum accumulated over time, whereas the large-scale jet is shaped by the collimating effect of the ambient medium at larger scales~\cite{Rohoza:2023egi}. The right panel of Fig.~\ref{fig:innerjet-fitting} illustrates the 3D structure of the resultant jet. Note that, at scales even closer to the SMBH, the structure of both the disk and the jet may align with the BH spin due to the magneto-spin alignment mechanism~\cite{mcKinney2013}. The extent of this alignment depends on the strength of the magnetic flux~\cite{mcKinney2013}. Given that the precessing disk is constrained to be compact and the accretion state is not fully concluded in M87 (see Methods), this alignment configuration is expected to be more compact. Therefore, we have ignored it in our analysis.

Observations of the jet structure and dynamics at various scales, especially at sub-mas scales, are crucial for fully understanding the detailed structure of the jet. In addition to the jet precession observed at $0.7 \sim 3$ mas~\cite{cui2023}, the sub-0.1-mas jet structure~\cite{Lu:2023bbn} provides notable constraints. In particular, Lu et al. (2023) reported an unexpectedly large jet width at $\lesssim 0.1$ mas~\cite{Lu:2023bbn}, exceeding the expected BZ jet~\cite{blandford1977} envelope and suggesting the need for an additional explanation in an untilted system. This discrepancy can be addressed within a precessing jet framework: the larger projected jet width with a smaller viewing angle in the inner regions, which arises naturally when the BH-driven jet is in a phase nearly coplanar with the line of sight (LOS) and the precession axis. The position angle of the jet at $\lesssim 0.1$ mas aligns roughly with that at larger scales, further supporting the interpretation that the inner jet precession was in such a phase at the time of observation. For simplicity, we assume that the inner jet was exactly coplanar with the LOS and the precession axis in 2018.

\begin{figure}
    \centering
    \includegraphics[width=0.7\linewidth]{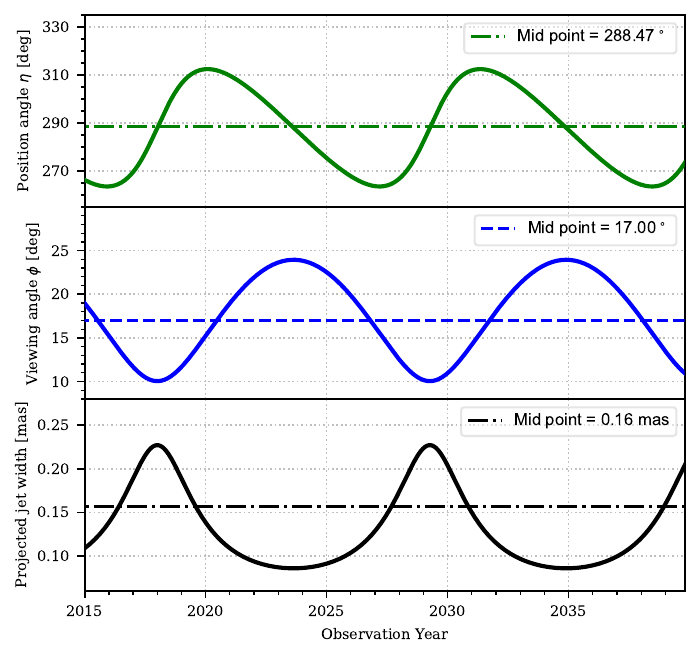}
    \caption{\textbf{Prediction of the evolution of jet features in the curved jet model.} The position angle ($\eta$), viewing angle ($\phi$), and projected jet width as functions of time are shown for the curved jet model, derived using best-fit parameters at a projected core separation of $r = 0.02\,\rm mas$.}
    \label{fig:prediction}
\end{figure}

To demonstrate that the curved jet model, inspired by the observed jet precession, effectively addresses the reported anomaly in the inner jet width, we analyze the projected jet width profile and compare it with the measurements from Lu et al. (2023)~\cite{Lu:2023bbn}.
The shape of an uncurved jet is modeled by a nearly parabolic cone along the $Z$-axis, 
\begin{equation}\label{eq:jet-shape-parabolic-main}
    W = 2C \rho^{\alpha}\,,
\end{equation}
where $W$ represents the intrinsic jet width. The parameters $C=0.19$ and $\alpha=0.6$ are determined by matching the predicted profile of projected jet width for the uncurved jet case with $a=0.9$~\cite{nakamura2018, Lu:2023bbn}. The jet is then curved according to the parameterized precession angle (Eq.~\eqref{eq:precession-angle-parameterization}) and projected onto the plane perpendicular to the LOS to derive the predicted jet width as a function of the measured jet axial distance (see Methods). To account for the precession observation by Cui et al. (2023)~\cite{cui2023}, a prior on the precession angle of $(1.25\pm0.18)^\circ$ at $1.8$ mas is incorporated into the analysis. In the left panel of Fig.~\ref{fig:innerjet-fitting}, we compare the predicted jet width profile with observations from Lu et al. (2023)~\cite{Lu:2023bbn}. The curved jet model predicts a wider jet width (solid curve) than the collimated jet model (dashed curve) at distances $\lesssim 0.1$ mas, while maintaining similar jet widths at larger scales. The prediction of the best-fit precessing curved jet model with $\psi_{\rm in}=11.1^\circ$, $\rho_{\rm t}=100\,r_{\rm g}$ and $\Delta=0.40$ aligns closely with the observed data (red dots).

Continuous monitoring of the jet in the region close to the SMBH is crucial to test the above hypothesis. As the jet precesses, both its position angle and viewing angle vary, especially at the sub-0.1-mas scales. The top and middle panels of Fig.~\ref{fig:prediction} illustrate the time evolution of these angles on a projected scale of $0.02$ mas. Importantly, the larger precession angle in the inner region causes the viewing angle to decrease when the jet approaches LOS. This results in an asymmetrical time variation curve for the position angle, which exhibits a skewness. The direction of this skewness is determined by the sense of precession and, therefore, by the direction of the SMBH spin.
As mentioned above, the SMBH spin almost aligns with the large-scale jet. With this, it can be inferred that the SMBH spin is oriented away from the Earth assuming a finite spin value~\cite{eht52019}. Thus, the skewness of the position angle offers a promising test of the SMBH’s spin direction. While the time variation of the viewing angle is more challenging to observe directly, it leads to changes in the jet width, as shown in the bottom panel of Fig.~\ref{fig:prediction}. Long-term observations resolving sub-mas scales~\cite{hada2016gmva, kim2018gmva, Lu:2023bbn} are promising for detecting variations in both the position angle and jet width in this region.

The compact precessing disk suggested in this work challenges models with strong magnetization, as the size of the precessing disk correlates with magnetic field strength~\cite{fragile2007, Porth_2019, chatterjee2022, Tchekhovskoy2015, Gupta:2024shd}. Together with that EHT observations disfavor weakly magnetized disk models~\cite{eht52019, eht82021, eht92023}, these suggest that the M87 accretion disk likely resides in an intermediate magnetization regime or some more subtle magnetic field configurations. Ongoing GRMHD simulations, particularly those that systematically explore different magnetic field configurations, the compact disk's surrounding environment, and disk tearing conditions, are essential for establishing the physical properties of tilted accretion disk~\cite{fragile2024}.
The curvature of the M87 jet revealed in this study reinforces the need for multi-wavelength observations spanning diverse spatial and temporal scales to resolve finer jet structures with advanced imaging techniques~\cite{janssen2019, chael2019, tiede2022, tazaki2023, kim2024a} and track their evolution over time. Combined with theoretical efforts, these observations are essential for unveiling the dynamics of the surrounding materials and the processes driving jet formation in the extreme environment near SMBH.

\begin{methods}
\section*{The precession of a coherently precessing disk}\label{sec:equ1}
\subsection*{The precession of a ring of test particles}
We start with the LT precession of a ring of test particles at a fixed radius around a spinning SMBH. The orbital angular frequency ($\Omega_{\rm orb}$) of such a ring near the equatorial plane is given by~\cite{kato1990, Lubow+2002, Franchini:2015wna}:
\begin{equation}\label{eq:orbital-angular-frequency}
    \Omega_{\rm orb}(r)=\frac{c^3}{GM}\frac{1}{(r/r_{\rm g})^{\nicefrac{3}{2}}+a}\,.
\end{equation}
This relation accounts for the relativistic effects of frame-dragging due to the SMBH's spin.
For orbits tilted slightly out of the equatorial plane, the deviation from the plane oscillates with a vertical frequency $\Omega_{z}$, which differs from $\Omega_{\rm orb}$. This difference arises from the relativistic frame-dragging effects. The relationship between $\Omega_{\rm orb}$ and $\Omega_{z}$ satisfies~\cite{Franchini:2015wna, lodato2013}:
\begin{equation}\label{eq:vertical-frequency}
    \frac{\Omega_{\rm orb}^2-\Omega_{z}^2}{\Omega_{\rm orb}^2} = 4a\left(\frac{r_{\rm g}}{r}\right)^{\frac{3}{2}}\left[1-\frac{3a}{4}\left(\frac{r_{\rm g}}{r}\right)^{\frac{1}{2}}\right]\,.
\end{equation}
The precession of the tilted orbital plane arises from the difference between the orbital ($\Omega_{\rm orb}$) and vertical ($\Omega_{z}$) frequencies, that is $\Omega_{\textsc{lt}}(r)=\Omega_{\rm orb}-\Omega_{z}$. From Eq.\,\eqref{eq:vertical-frequency}, we can solve for $\Omega_{\textsc{lt}}$ which gives (we denote the right-hand side of Eq.\,\eqref{eq:vertical-frequency} as $\delta$)
\begin{equation}\label{eq:LT-angular-frequency}
\begin{split}
\Omega_{\textsc{lt}}(r) &= \Omega_{\rm orb}\big(1-\sqrt{1-\delta}\big)\\
&\simeq\frac{1}{2}\Omega_{\rm orb}(\delta+\frac{1}{4}\delta^2)\\
&\simeq\frac{2aG^2M^2}{c^3r^3}\left[1-\frac{3a}{4}\left(\frac{r_{\rm g}}{r}\right)^{\frac{1}{2}}\right]\,.
\end{split}
\end{equation}
Note that the correction terms in the parentheses with order $\mathcal{O}(r_{\rm g}/r)^{\frac{3}{2}}$ have been canceled out upon expansion and terms with order higher than $\mathcal{O}(r_{\rm g}/r)^{\frac{3}{2}}$ have been ignored. The precession period $T_{\rm prec} = 2\pi/\Omega_{\textsc{lt}}$, shown in Eq.\,\eqref{eq:LT-precession-effective-higher-order}, is highly sensitive to the disk’s size.

\subsection*{An extended coherently precessing disk}

The precessing disk can be approximated as a torus, characterized by an inner radius $r_{\rm in}$ and an outer radius $r_{\rm o}$. Such a coherently precessing structure has been demonstrated in recent simulations~\cite{fragile2007, lodato2013, cui2023}. The precessing disk constitutes only a portion of the accretion disk, rather than its entirety. It is important to note that $r_{\rm in}$ and $r_{\rm o}$ should not be interpreted as sharply defined edges but rather as effective radii describing the approximate extent of the precessing disk. They are determined by the interplay between the frame-dragging effects of the spinning SMBH and the internal interactions within the disk. As the torus precesses as a whole, its inner (outer) regions precess more slowly (rapidly) compared to how they would behave as independent rings. This dynamic behavior necessitates defining an effective radius $r_{\textsc{lt}}$, which depends on the precessing disk’s mass distribution and serves to characterize its overall response to the frame-dragging effects of a spinning SMBH.

To describe a coherently precessing disk with finite radial extent and thickness, we define an angular momentum-averaged precession frequency as~\cite{fragile2007, nixon2012}
\begin{equation}\label{eq:effective-LT-frequency}
    \Omega_{\rm prec} \equiv\frac{\int_{r_{\rm in}}^{r_{\rm o}} \Omega_{\textsc{lt}}(r)l(r)\times2\pi rdr}{\int_{r_{\rm in}}^{r_{\rm o}}l(r)\times2\pi rdr}\,, 
\end{equation}
where $l(r)$ is the angular momentum density. The effective LT radius, $r_{\textsc{lt}}$, is then defined by the condition,
\begin{equation}\label{eq:effective-lt-radius-defintion}
    \Omega_{\rm prec}\equiv \Omega_{\textsc{lt}}(r_{\textsc{lt}})\,.
\end{equation}
The surface density profile of the precessing disk (integrated over the altitude direction) is assumed to follow $\sigma(r)\propto r^{-\zeta}$. Since the region dominating the integrals in Eq.\,\eqref{eq:effective-LT-frequency} lies far from the BH, we follow~\cite{fragile2007} and approximate $\Omega_{\textsc{lt}}(r)$ with its leading order in the radius dependence. Under the assumption of a Keplerian orbital velocity $v\propto r^{-1/2}$, the leading-order precession frequency is:
\begin{equation}
    \Omega_{\rm prec}^{\rm leading}=\frac{2aG^2M^2}{c^3r_{\textsc{lt}}^3}=\frac{2aG^2M^2}{c^3}\frac{1}{r_{\rm o}^{\frac{5-2\zeta}{2}}r_{\rm in}^{\frac{1+2\zeta}{2}}}\frac{5-2\zeta}{1+2\zeta}\frac{1-(r_{\rm in}/r_{\rm o})^{\frac{1}{2}+\zeta}}{1-(r_{\rm in}/r_{\rm o})^{\frac{5}{2}-\zeta}}\,.
\end{equation}
This provides a direct link between the disk's precession and its radial structure. The effective LT radius ($r_{\textsc{lt}}$) arising from this relationship, is given by Eq.\,\eqref{eq:LT-radius-fragile}.

\textit{Note}. 
The recent study~\cite{Wei:2024cti} also examines the implications of M87 jet precession on the properties of the SMBH spin and accretion disk. Their Fig.~8 resembles our Fig.~\ref{fig:period-a-rlt} if we equate our $r_{\textsc{lt}}$ with their warp radius. However, while they assume that the jet's motion follows a ring of the disk at the warp radius that precesses independently from the rest of the disk, our interpretation is that a torus with a certain size and the jet precess coherently as demonstrated in recent simulations.
In terms of disk morphology, the warp radius serves a role similar to our inner radius of the coherently precessing disk bulk. We note that $r_{\textsc{lt}}$ is an effective radius dependent on the mass distribution within the precessing disk. Given these distinctions, we conclude that the coherently precessing disk is compact and we further investigate the extended disk size. Our lower bound of $a$ is the same as that given in~\cite{Wei:2024cti}, since in that case, the coherently precessing disk is confined to a ring with $r=r_{\textsc{isco}}$.

\subsection*{The dependence on the tilt angle}
Thus far, our analysis has focused on LT precession near the equatorial plane, and Eq.\,\eqref{eq:LT-precession-effective-higher-order} does not account for dependence on the tilt angle.
The prograde and retrograde scenarios considered in this work represent the two extreme cases. The constraints on the $a$-$r_{\textsc{lt}}$ parameter space and disk morphology, as presented in Figs. \ref{fig:period-a-rlt} and \ref{fig:constraint-details}, encapsulate the results for these limiting scenarios. For a general case with an arbitrary tilt angle, the constraints are expected to lie between these two extremes.

The LT precession of a single ring with an arbitrary tilt angle can be computed using the framework outlined in~\cite{Wei:2024cti}, where different tilt angles correspond to different Carter constants. However, the tilt angle is constrained to $\lesssim 60^\circ$~\cite{wielgus2020}. From our curved jet model, the inner precession angle of the jet is constrained to approximately $11^\circ$, suggesting a correspondingly small tilt angle for the accretion disk. This supports our approach of focusing on disk precession near the equatorial plane.

\subsection*{Remarks on potential tests on modified gravity in the strong-field region}

The observations of jet precession potentially present a unique opportunity to test modified gravity theories in the strong-field regime. For instance, according to the no-hair theorem, all higher multipole moments (quadrupole and beyond) of a BH are determined solely by its mass and angular momentum~\cite{Geroch:1970cd}. A deviation from the predicted quadrupole moment under the no-hair theorem would lead to a different LT precession period~\cite{Glampedakis:2005cf, Yagi:2016jml} for a given effective LT radius. Another example is Chern-Simons gravity, which involves gravitational parity violation. In this framework, LT precession is modified due to the correction in the gravitomagnetic sector~\cite{Alexander:2009tp}, also resulting in a precession period for a given effective radius. However, testing such deviations is challenging due to considerable uncertainties in the properties of the accretion disk. Variations in disk geometry, viscosity, and magnetization can introduce degeneracies that either mimic or obscure deviations from the Kerr solution. Consequently, disentangling potential signals of modified gravity from these astrophysical effects remains a complex but promising endeavor.

\section*{Curved jet and the inner jet projected width}\label{sec:curved-jet-projection}
As demonstrated in Section~\ref{sec:innermost-jet}, the inner jet may have a larger precession angle, which can explain the unexpectedly large measured jet width at $\lesssim0.1$ mas reported in Lu et al. (2023)~\cite{Lu:2023bbn}. Here, we highlight the technical details of curved jet modeling and parameter inference. We assume that the edge shape of the 3D jet can be parameterized by a nearly parabolic function given in Eq.\,\eqref{eq:jet-shape-parabolic-main}. Both $W$ and $\rho$ are measured in mas, with a conversion factor of $1\,\text{mas} = 250 \,r_{\rm g}$ for M87.

We begin by reproducing the theoretical prediction of the jet width as a function of the measured axial distance for $a=0.9$. The coordinate system is defined with the $X$--$Y$ plane perpendicular to LOS and the $Z$ axis pointing towards the observer. We construct a parabolic cone along the $Z$ direction, tilted by $17^\circ$ in the $Z$--$Y$ plane. The projection of this cone onto the $X$--$Y$ plane represents the 2D jet observed. The right panel of Fig.~\ref{fig:innerjet-fitting} illustrates the schematic of a tilted jet and its projection on the $X$--$Y$ plane (the depicted case of a curved jet will be discussed later). In this projection, the width in the $X$-direction corresponds to the observed jet width, and the $Y$-coordinate corresponds to the observed axial distance. By matching the predicted jet width as a function of the axial distance, we determine the parameters $C=0.19$ and $\alpha=0.6$. In \SUPFIG{supfig:InnerJet-suppl}, the left panel presents a 3D view of the curved jet from an edge-on view. The middle panel illustrates the 2D structure of the jet axis, while the right panel displays the profiles of the precession angle and the viewing angle.

\setcounter{figure}{0}
\captionsetup[figure]{name={\bf Supplementary Figure}}
\begin{figure}
    \centering
    \includegraphics[width=0.99\linewidth]{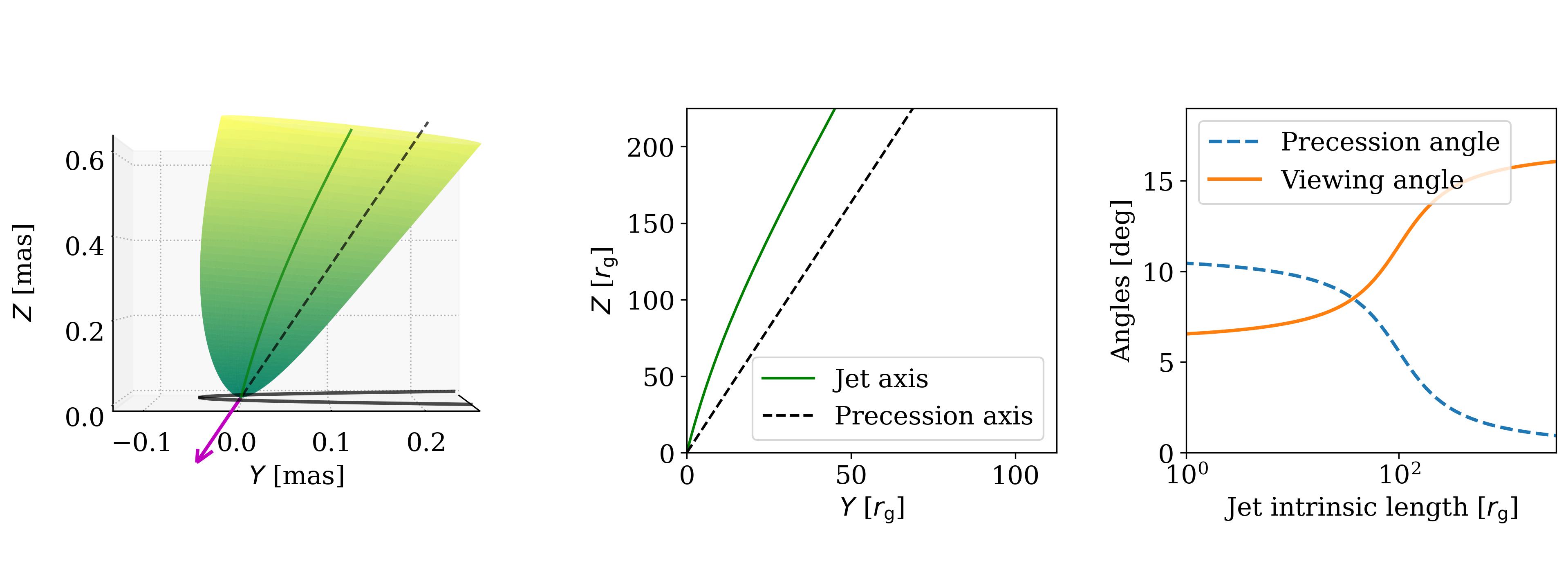}
    \caption{\textit{Left}: The 3D curved jet viewed edge-on, with symbols corresponding to those in the right panel of Fig.~\ref{fig:innerjet-fitting}. The $Y$-axis represents the projected distance from the core, while the $Z$-axis means the distance from the core along LOS (positive means pointing towards us). \textit{Middle}: The jet axis projected onto a 2D plane, providing a clearer view of its curvature. \textit{Right}: Profiles of the precession angle and the viewing angle. }
    \label{supfig:InnerJet-suppl}
\end{figure}

The scenario described above is the standard case assumed in the analysis of Lu et al. (2023)~\cite{Lu:2023bbn}. In contrast, our curved jet model proposes that, rather than forming a collimated cone, the angle between the jet and its precession axis (referred to as the precession angle, $\psi$) decreases from a finite value near the core to zero at large distances as parameterized by Eq.\,\eqref{eq:precession-angle-parameterization}. 
Motivated by the observation that the jet PA at scales $\lesssim 0.1$ mas is similar to that at larger scales, the precession of the inner jet is likely in a phase in which the inner jet, the large-scale jet, and the LOS are nearly coplanar. Consequently, the jet viewing angle as a function of $\rho$ is given by
\begin{equation}\label{eq:}
    \phi(\rho) = 17^\circ - \psi(\rho)\,,
\end{equation}
assuming $\psi<17^\circ$. The value of $17^\circ$ is adopted from the literature~\cite{mertens2016, walker2018, cui2023}. Similar to the standard case, we project the curved jet onto the $X$--$Y$ to derive the observed jet width as a function of the measured axial distance. 

The projected jet width is determined analytically. The equation for the surface of the curved jet is derived as follows. First, each point $A$ on the uncurved jet axis is mapped to a corresponding point $A'$ on the curved axis, maintaining the same intrinsic length $\rho$. We denote the $Y$- and $Z$-coordinate of $A'$ as $\rho_{\rm y}$ and $\rho_{\rm z}$, respectively. The uncurved jet cone is divided into rings of constant $z$ (corresponding to the length along the jet axis). Each ring is first shifted to the origin by a vector $(0,0,-\rho)$, rotated by an angle $\phi(\rho)$, and then shifted again by a vector $(0, \rho_{\rm y},\rho_{\rm z})$. The resulting ring satisfies the following equations
\begin{align}
    &x^2+\frac{(y-\rho_{\rm y})^2}{\cos^2\phi}-C^2\rho^{2\alpha}=0\,, \label{eq:new-ring-ellipse} \\
    &z = \rho_{\rm z} -\frac{\sin\phi}{\cos\phi}(y-\rho_{\rm y})\,.
\end{align}
These transformed rings collectively form the surface of the curved jet. The observed jet width at a given measured jet axial distance $y$ is twice the maximum $X$-coordinate on the curved jet's surface. From Eq.\,\eqref{eq:new-ring-ellipse}, the value of $\rho$ that maximizes $x$ for a given $y$ satisfies
\begin{equation}\label{eq:eq-for-rho}
    (y-\rho_{\rm y})\frac{\sin\phi}{\cos^2\phi}+\alpha C^2\rho^{2\alpha-1}-(y-\rho_{\rm y})^2\frac{\sin\phi}{\cos^3\phi}\frac{d\phi}{d\rho}=0\,.
\end{equation}
The observed jet width can then be calculated as:
\begin{equation}
    W(y)=2\times\sqrt{C^2\rho^{2\alpha}-\frac{(y-\rho_{\rm y})^2}{\cos^2\phi} }\,.
\end{equation}
Examples of $W(y)$ for both a curved jet and an uncurved jet are shown in the left panel of Fig.~\ref{fig:innerjet-fitting}, represented by the solid and dashed curves, respectively. The observed jet width is larger for a curved jet compared to a collimated one. This occurs because, with smaller viewing angles at the inner region, the same measured axial distance $y$ corresponds to a longer intrinsic jet length $\rho$, resulting in a wider jet. 

We performed an MCMC analysis of the curved jet model, comparing it with the jet width profile measured by Lu et al. (2023)~\cite{Lu:2023bbn}. The results are presented in \SUPFIG{supfig:MCMC}. Current jet width observations do not uniquely constrain the transition parameters, as there is a degeneracy between the transition position $\rho_{\rm t}$ and the inner precession angle $\psi_{\rm in}$. A smaller $\rho_{\rm t}$, combined with a larger $\psi_{\rm in}$, can yield the same observed jet width profile. We introduced a prior of $\rho_{\rm t} > 100\,r_{\rm g}$ in the MCMC analysis, motivated by the fact the inner jet remains perpendicular to the precessing disk at least out to some intermediate scale~\cite{cui2023,liska2018,mcKinney2013}, although releasing this prior does not affect the goodness of fit or our main conclusion.

\begin{figure}
    \centering
    \includegraphics[width=0.7\linewidth]{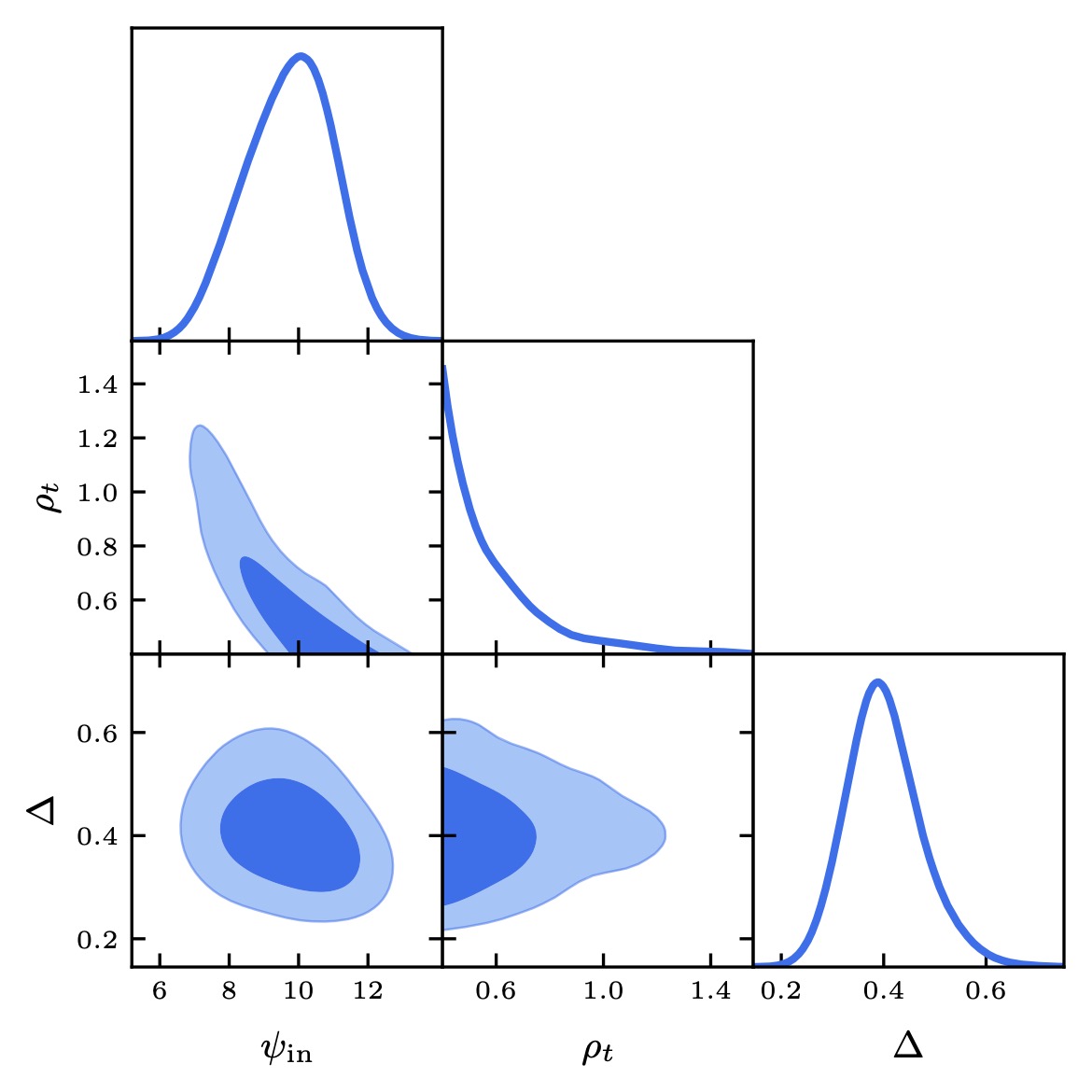}
    \caption{Results of the MCMC analysis for the curved jet model, compared with the projected jet width profile measured by Lu et al. (2023)~\cite{Lu:2023bbn}. The transition length $\rho_{\rm t}$ is in the unit of mas, which can be converted to the physical length with 1 mas = 250 $r_{\rm g}$ for M87.}
    \label{supfig:MCMC}
\end{figure}

To predict the evolution of the position angle, viewing angle, and jet width at a projected distance of $0.02$ mas from the core, we first note that this scale is closer to the core than the transition region. Therefore, we approximate the jet at this scale using a fixed precession angle. Any minor discrepancies from this approximation are addressed by introducing an effective precession angle, which matches the measured jet width in 2018. At different times, the position and viewing angles are calculated following the method in Cui et al. (2023)~\cite{cui2023}, while the jet width is determined based on the previously outlined method, taking into account the evolution of the viewing angle.

The applicability of parabolic jet models for M87 within the Bondi radius \cite{asada2012, nakamura2018, hada2016gmva, kim2018gmva, chatterjee2019}, which is a characteristic feature of the BH-driven jet \cite{nakamura2018}. To highlight the role of jet curvature, we have adopted a parabolic model for the uncurved jet, consistent with the approach used in \cite{Lu:2023bbn} which is based on the simulation performed in \cite{pu2016}. As noted in \cite{liska2019}, systems with geometrically thin accretion disks can produce jets with larger opening angles than the ones produced by thick disks beyond $\sim3\,r_{\rm g}$ due to reduced pressure support. While this mechanism may explain the broader inner jet width observed in M87, it remains unclear—and likely challenging—to consistently account for the narrower jet width observed at larger scales. Furthermore, the thin disk assumption in \cite{liska2019} conflicts with the thick disk geometry for M87 as inferred from its hot, low-accretion-rate flow \cite{hada2024}. One might argue that both the thin- and thick-disk cases have a similar wide opening angle near the horizon ($\lesssim 3\,r_{\rm g}$). However, it is important to clarify that the region of the jet near the horizon contributes negligibly to the observed jet width profile due to projection effects, as we explain below. As shown in Fig.~\ref{fig:innerjet-fitting} and discussed in the Methods section, the observed jet width is the projection of a three-dimensional structure. In the left panel of Fig.~\ref{fig:innerjet-fitting}, we see that the observed jet width remains finite at approximately 0.2 mas. Even at the projected distance of zero, the observed width roughly corresponds to the intrinsic jet width at $85\,r_{\rm g}$. The jet structure at $\sim 3\,r_{\rm g}$ has a negligible contribution to the observed width profile due to projection effects, as it is obscured by the larger-scale jet regions. This effect is clearly visualized in the right panel of Fig.~\ref{fig:innerjet-fitting}. Instead, the projection of a curved jet plays a key role in shaping the observed width profile, as illustrated in the right panel of Fig.~\ref{fig:innerjet-fitting} and the left panel of \SUPFIG{supfig:InnerJet-suppl}.

\section*{Current status of GRMHD simulations}
\subsection*{The magnetization state of accretion disk}
The magnetization state of the accretion disk around SMBHs remains a central topic in understanding the accretion process and jet formation. Two primary states are commonly discussed: the Magnetically Arrested Disk (MAD) state~\cite{igumenshchev2003, narayan2003, tchek2011, mcKinney2012, narayan2022, chatterjee2022} and the Standard and Normal Evolution (SANE) state~\cite{DeVilliers2003, gammie2003, narayan2012, Porth_2019}. The normalized magnetic flux is defined as 
$\phi_{\textsc{bh}}\equiv\Phi_{\textsc{bh}}/\sqrt{\dot{M}_{\textsc{bh}}r_{\rm g}^2c}$, where $\Phi_{\textsc{bh}}$ is the magnetic flux threading BH with the magnetic field in Gaussian units. This $\phi_{\textsc{bh}}$ is typically $\gtrsim 50$ for MAD state and $\lesssim 5$ for SANE state. The MAD state emerges when the magnetic flux threading the disk reaches a saturation level, choking accretion and causing episodic flux eruptions, while the SANE state features weaker, turbulent magnetic fields that permit smoother and more continuous accretion flows~\cite{chatterjee2022}.

For the SMBH in M87, recent EHT observations provide crucial insights into its magnetization states. The EHT 2017 total intensity image of the photon ring surrounding the BH is consistent with parts of both MAD and SANE models~\cite{eht52019}. Further constraints come from jet power. The MAD state more naturally explains M87's powerful jet exceeding $10^{42}\,\rm erg\,s^{-1}$, compared to SANE models~\cite{eht52019, eht82021, eht92023}. Interestingly, the jet power alone cannot rule out SANE models, as varying the $\phi_{\textsc{bh}}$ parameter in SANE systems can meet the required jet power~\cite{sadowski2013, aktar2024}, especially when driven by processes such as spin-enhanced magnetic coupling~\cite{eht52019}. However, the inclusion of polarization maps disfavors the SANE state, as suggested by the observed high fractional polarization and coherent magnetic field structure~\cite{eht82021, eht92023}. Note that there are uncertainties in the EHT constraints from limited simulation diversity, assumptions of an aligned disk-BH system, and potential magnetic flux variability or possible transitions to MAD state in SANE cases~\cite{yao2021, Tsunetoe:2022ktx}. Additionally, it is important to note that classifying the disk’s magnetic state solely based on the magnetic flux threading the black hole may not be appropriate. Even within the SANE or MAD states, there can be a broad range of magnetic field strengths within the disk. The magnetic field strength and configuration in the accretion disk are subjects of ongoing research \cite{fragile2024}.

\subsection*{Challenges and potential resolutions}
Extensive efforts have been devoted to numerical simulations on tilted-accretion-disk systems~\cite{fragile2005, fragile2007, fragile2009, mcKinney2013, sadowski2013, liska2018, liska2019, white2019, ressler2021, liska2021, Chatterjee_etal:2023, ressler2023, ressler2024MAD}, which allow us to have some general understandings on the dynamics of these systems. However, most GRMHD simulations rely on simplified initial conditions---starting with small, finite tori of gas threaded by magnetic fields---primarily for computational convenience~\cite{fragile2024}. While these setups serve as useful numerical experiments, they may not accurately reflect the physical conditions of BH accretion flows in nature. In reality, accretion likely occurs from large radii, with gas circularizing far from the black hole. If the entire disk remains a unity, such a setup tends to cause the outer regions of the simulated disk to expand in order to conserve angular momentum, which decisively affects the precession dynamics.

In reality, the disk structure may be more complex. A more physically realistic configuration could involve an inner, hot, and thick disk, fed by a larger, thinner accretion flow (a ``truncated disk''), which prevents the inner torus from expanding~\cite{bollimpalli2023, bollimpalli2024}. In this scenario, the inner disk would still undergo precession, but at a reduced rate of approximately $5\%$ than an isolated and continuous disk. It remains uncertain whether this also applies to M87, given that its accretion rate is substantially lower than that used in these simulations. If the truncated disk configuration holds for M87, the reduced precession rate (assumed to be also $5\%$) would imply an even more compact effective LT radius  ($\lesssim6\,r_{\rm g}$). Another intriguing possibility is a compact disk that has been torn off from a larger disk, an area of ongoing research that explores the dynamics, stability, and formation mechanisms of such systems, which may involve complex interactions between disk instabilities, magnetic fields, and external perturbations~\cite{fragile2024,dogan2018}.

Reproducing LT precession is more challenging if the strong magnetic field is additionally considered~\cite{fragile2024,Chatterjee_etal:2023,mcKinney2013}. Currently, it is found that in the presence of a strong magnetic field, the disk tends to align with the BH spin (``magneto-spin alignment'' mechanism \cite{mcKinney2013}). The size of this region is sensitive to the strength of the magnetic field \cite{mcKinney2013, Porth_2019,chatterjee2022, Tchekhovskoy2015, Gupta:2024shd}. For MAD-state simulations in~\cite{liska2018, Chatterjee_etal:2023}, the alignment size exceeds $>100\,r_{\rm g}$, making the period of any possible LT precession to exceed $10^6\,r_{\rm g}/c$ according to Eq.\,\eqref{eq:LT-precession-effective-higher-order}, which would be too long for SMBH such as that in M87.

One way to mitigate the strong alignment effect mentioned above might involve considering intermediate magnetic field strengths within the accretion disk. Although the semi-MAD state was not explicitly analyzed in~\cite{eht52019,eht92023}, it has been demonstrated to effectively account for the EHT polarization maps~\cite{Tsunetoe:2022ktx}. While most numerical works focused on either the SANE or MAD regimes, the recent simulation~\cite{cui2023} of a tilted semi-MAD state accretion disk, with a magnetic flux strength ($\phi_{\textsc{bh}}\simeq17$) consistent with those studied in~\cite{Tsunetoe:2022ktx}, successfully reproduce LT precession of the disk. Notably, the reproduced precession period of approximately 11 years~\cite{cui2023} is consistent with observations. However, while this simulation might reduce the alignment size, it could be under-resolved. If the resolution were high enough to capture turbulence effects, the disk may expand, reverting to a large isolated accretion disk and strikingly lengthening the LT precession period~\cite{liska2018}. Therefore, it may be necessary to incorporate additional physics, such as a truncated disk configuration, to maintain a compact disk size.

Additionally, as LT procession is one of the most promising explanations for the luminosity variation for black hole X-ray binaries, there have been concerns on how the system can achieve a MAD state; or that present simulations are insufficient to conclude ``MADs do not precess''~\cite{Fragile:2023qpz}.

In summary, although simulating LT precession in accretion disks within GRMHD frameworks remains challenging, further investigation into the physics of tilted accretion disks is essential. This includes exploring magnetic configurations, disk structure, disk tearing, and the influence of the surrounding environment, all of which are crucial for a more comprehensive understanding of these systems.

\section*{Alternative scenarios for the M87 structure variation}\label{sec:alternative}

We attribute the observed variation in the jet position angle to LT precession of the accretion disk, a hypothesis supported by the coherent position angle variation reported in \cite{cui2023} and strengthened by the successful explanation for the recently reported sub-0.1 mas jet width anomaly. While we acknowledge the current challenge in reproducing LT precession of the accretion disk, particularly in the presence of strong magnetic fields, we will discuss some other physical processes that may potentially explain the position angle variation and/or the inner jet width profile. Below, we briefly examine these alternative scenarios.

\subsection*{Binary black hole}

A widely considered explanation for periodic disk/jet behavior involves binary black hole (BBH) systems~\cite{begelman1980, valtonen2008}, with three potential mechanisms contributing to the periodicity of the jet position angle variation. First, spin-orbit coupling can induce precession of the primary BH’s spin~\cite{ressler2024MAD, ressler2025BBH}. Second, the secondary BH’s frame-dragging effect can cause the tilted accretion disk around the primary BH to precess~\cite{Shen:2024gwb}. Third, the secondary BH may interact with the disk/jet system, e.g., by repeatedly piercing the accretion disk~\cite{begelman1980, valtonen2008} or somehow directly interacting with the jet.

However, these BBH scenarios face challenges in explaining the M87 system. High-accuracy astrometric VLBA 43 GHz observations have revealed an extremely stable astrometric position of the M87 radio core (with a scatter of only $12\,r_{\rm g}$) over several years~\cite{acciari2009}, which disfavors the presence of a secondary gravitational source near the primary BH. For the first two mechanisms, achieving the relatively short variation timescale of $\sim 11$ years requires a moderate secondary-to-primary mass ratio and a small BH separation (e.g., a mass ratio of $0.1$ and a separation of $<30\,r_{\rm g}$~\cite{ressler2025BBH}). This would lead to a BBH merger timescale much shorter than the cosmic timescale, making these explanations highly improbable. For the third mechanism, while the variation timescale is linked to the binary orbital period, the specific process by which the secondary BH affects the jet position angle remains unclear and requires further investigation.

\subsection*{Stochastic/quasi-periodic scenarios}
There are two types of instabilities mainly discussed in AGN jet studies, namely Kink instabilities~\cite{mizuno2009, mizuno2012} and Kelvin-Helmholtz (KH) instabilities~\cite{blandford1976, mizuno2007}. Kink instabilities, a class of MHD instabilities in the strongly magnetized regime, occur in jets dominated by toroidal magnetic fields. These instabilities can induce helical or oscillatory deformations, particularly in systems with strong magnetic fields, such as those associated with MADs. KH instabilities arise from the interaction between the jet and its surrounding medium in the weakly magnetized regime, generating turbulent structures and transverse oscillations at various scales. This type of instability has been proposed to explain the structural variations in the kpc-scale jet of M87~\cite{lobanov2003, pasetto2021} and pc-scale helical structure~\cite{nikonov2023}, where they produce complex, non-linear dynamics. KH instabilities can also manifest at smaller scales, potentially influencing the observed PA variation at mas scales. Walker et al. (2018)~\cite{walker2018} suggested that such instabilities might explain the quasi-periodic jet oscillations they observed. 

Several other scenarios may also lead to random variations in the jet’s position angle. For instance, the magnetic state of the accretion disk can be variable, which may cause some temporal changes in the jet’s width and orientation \cite{tsunetoe2024}. Similarly, magnetic flux eruptions can induce fluctuations in the tilt of the accretion disk, with an amplitude of 3 to 5 degrees \cite{Chatterjee_etal:2023}, potentially explaining the observed amplitude of the position angle variations reported by Cui et al. (2023). Interestingly, the time separation of between the largest variation is about 10 years. Additionally, the tilt angle of the disk can vary due to the random orientation of gas infall in a spherical inflow configuration \cite{Lalakos:2023ean}. However, these scenarios typically lead to non-coherent and stochastic jet behaviors, rather than the periodic PA variation seen in \cite{cui2023}. 

In the second part of this work, we demonstrated that a curved jet model can effectively explain the inner jet width profile. While the curved jet model is motivated by the jet precession hypothesis, the jet could also curve if launched from a warped, non-precessing disk~\cite{liska2018}. In this scenario, additional physical mechanisms would be required to explain the jet’s position angle variation, whose coherent features are difficult to account for without jet precession as pointed out above. Nevertheless, long-term monitoring of the inner jet width variations is crucial for distinguishing between these two possibilities.

Future yearly and long-term monitoring of the M87 jet across multiple frequencies (e.g., \cite{Lu:2023bbn, walker2018, Giovannini2023, Pushkarev2017, MWL2024}) will be crucial for further investigating the nature of the structural variations in the M87 jet’s position angle. Advancements in imaging techniques~\cite{akiyama2017, janssen2019, chael2019, tiede2022, tazaki2023, kim2024a, kim2024b} hold the potential to unveil finer jet structures and their evolution over time. Furthermore, the magnetic sheath surrounding the jet could be broader than the jet itself \cite{lisakov2021}. In that case, as the jet swings within the magnetic sheath, this motion could lead to associated sign changes in the observed rotation measure (RM). Therefore, long-term RM monitoring offers a potential method to distinguish whether the jet position variations are periodic or stochastic.

\end{methods}

\bmhead{Data availability}
No specific datasets were generated in this study. The jet width profile data presented in Fig.~3 are adopted from Lu et al. (2023)[15].

\bmhead{Code availability}
No custom codes or specialized software scripts were created for this study. Any calculations or procedures mentioned are fully described in the Methods section, and they can be reproduced using commonly available tools or standard software.

\bmhead{Corresponding author}
Correspondence and requests for materials should be addressed to Yuzhu Cui (yuzhu\_cui77@163.com) or Weikang Lin (weikanglin@ynu.edu.cn).

\bmhead{Acknowledgements}
We thank Kazuhiro Hada and Mareki Honma for their valuable discussion and comments, and Rusen Lu for sharing the measurements of the jet width from Lu et al. (2023). We would like to express our sincere gratitude to P. Chris Fragile for his thoughtful and detailed comments, suggestions, and explanations regarding the current numerical simulations, as well as for his insights into future directions in theory. We are grateful to the South-Western Institute For Astronomy Research of Yunnan University for their support during Y. C.'s visit. Open access funding is provided by the Key Program of National Nature Science Foundation of China under grant No. 12033001. W. L. acknowledges that this work is supported by the ``Science \& Technology Champion Project'' (202005AB160002) and the ``Top Team Project'' (202305AT350002), both funded by the ``Yunnan Revitalization''. W. L. is also supported by the ``Yunnan General Grant'' (202401AT070489). Y. C. is supported by the Natural Science Foundation of China (grant 12303021) and the China Postdoctoral Science Foundation (No. 2024T170845).

\bmhead{Author contributions}
Both authors jointly contributed to all parts of this work including the research design, data analysis, and manuscript preparation. 

\bmhead{Ethics declarations}
The authors declare no competing interests.

\bibliography{ref}

\providecommand{\noopsort}[1]{}\providecommand{\singleletter}[1]{#1}%
\begin{thebibliography}{100}
\expandafter\ifx\csname url\endcsname\relax
  \def\url#1{\burl{#1}}\fi
\expandafter\ifx\csname urlprefix\endcsname\relax\def\urlprefix{URL }\fi
\providecommand{\bibinfo}[2]{#2}
\providecommand{\eprint}[2][]{\url{#2}}
\providecommand{\doi}[1]{\url{https://doi.org/#1}}
\bibcommenthead

\bibitem{curtis1918}
\bibinfo{author}{{Curtis}, H.~D.}
\newblock \bibinfo{title}{{Descriptions of 762 Nebulae and Clusters
  Photographed with the Crossley Reflector}}.
\newblock \emph{\bibinfo{journal}{Publications of Lick Observatory}}
  \textbf{\bibinfo{volume}{13}}, \bibinfo{pages}{9--42} (\bibinfo{year}{1918}).

\bibitem{igumenshchev2003}
\bibinfo{author}{{Igumenshchev}, I.~V.}, \bibinfo{author}{{Narayan}, R.} \&
  \bibinfo{author}{{Abramowicz}, M.~A.}
\newblock \bibinfo{title}{{Three-dimensional Magnetohydrodynamic Simulations of
  Radiatively Inefficient Accretion Flows}}.
\newblock \emph{\bibinfo{journal}{\apj}} \textbf{\bibinfo{volume}{592}},
  \bibinfo{pages}{1042--1059} (\bibinfo{year}{2003}).

\bibitem{yuan2014}
\bibinfo{author}{{Yuan}, F.} \& \bibinfo{author}{{Narayan}, R.}
\newblock \bibinfo{title}{{Hot Accretion Flows Around Black Holes}}.
\newblock \emph{\bibinfo{journal}{\araa}} \textbf{\bibinfo{volume}{52}},
  \bibinfo{pages}{529--588} (\bibinfo{year}{2014}).

\bibitem{eht52019}
\bibinfo{author}{{Event Horizon Telescope Collaboration}}.
\newblock \bibinfo{title}{{First M87 Event Horizon Telescope Results. V.
  Physical Origin of the Asymmetric Ring}}.
\newblock \emph{\bibinfo{journal}{\apjl}} \textbf{\bibinfo{volume}{875}},
  \bibinfo{pages}{L5} (\bibinfo{year}{2019}).

\bibitem{eht82021}
\bibinfo{author}{{Event Horizon Telescope Collaboration}}.
\newblock \bibinfo{title}{{First M87 Event Horizon Telescope Results. VIII.
  Magnetic Field Structure near The Event Horizon}}.
\newblock \emph{\bibinfo{journal}{\apjl}} \textbf{\bibinfo{volume}{910}},
  \bibinfo{pages}{L13} (\bibinfo{year}{2021}).

\bibitem{reynolds2013}
\bibinfo{author}{{Reynolds}, C.~S.}
\newblock \bibinfo{title}{{The spin of supermassive black holes}}.
\newblock \emph{\bibinfo{journal}{\cqg}} \textbf{\bibinfo{volume}{30}},
  \bibinfo{pages}{244004} (\bibinfo{year}{2013}).

\bibitem{Risaliti2013}
\bibinfo{author}{{Risaliti}, G.} \emph{et~al.}
\newblock \bibinfo{title}{{A rapidly spinning supermassive black hole at the
  centre of NGC 1365}}.
\newblock \emph{\bibinfo{journal}{\nat}} \textbf{\bibinfo{volume}{494}},
  \bibinfo{pages}{449--451} (\bibinfo{year}{2013}).

\bibitem{McClintock2006}
\bibinfo{author}{{McClintock}, J.~E.} \emph{et~al.}
\newblock \bibinfo{title}{{The Spin of the Near-Extreme Kerr Black Hole GRS
  1915+105}}.
\newblock \emph{\bibinfo{journal}{\apj}} \textbf{\bibinfo{volume}{652}},
  \bibinfo{pages}{518--539} (\bibinfo{year}{2006}).

\bibitem{McClintock2014}
\bibinfo{author}{{McClintock}, J.~E.}, \bibinfo{author}{{Narayan}, R.} \&
  \bibinfo{author}{{Steiner}, J.~F.}
\newblock \bibinfo{title}{{Black Hole Spin via Continuum Fitting and the Role
  of Spin in Powering Transient Jets}}.
\newblock \emph{\bibinfo{journal}{\ssr}} \textbf{\bibinfo{volume}{183}},
  \bibinfo{pages}{295--322} (\bibinfo{year}{2014}).

\bibitem{eht12019}
\bibinfo{author}{{Event Horizon Telescope Collaboration}}.
\newblock \bibinfo{title}{{First M87 Event Horizon Telescope Results. I. The
  Shadow of the Supermassive Black Hole}}.
\newblock \emph{\bibinfo{journal}{\apjl}} \textbf{\bibinfo{volume}{875}},
  \bibinfo{pages}{L1} (\bibinfo{year}{2019}).

\bibitem{eht62019}
\bibinfo{author}{{Event Horizon Telescope Collaboration}}.
\newblock \bibinfo{title}{{First M87 Event Horizon Telescope Results. VI. The
  Shadow and Mass of the Central Black Hole}}.
\newblock \emph{\bibinfo{journal}{\apjl}} \textbf{\bibinfo{volume}{875}},
  \bibinfo{pages}{L6} (\bibinfo{year}{2019}).

\bibitem{blandford1977}
\bibinfo{author}{{Blandford}, R.~D.} \& \bibinfo{author}{{Znajek}, R.~L.}
\newblock \bibinfo{title}{{Electromagnetic extraction of energy from Kerr black
  holes.}}
\newblock \emph{\bibinfo{journal}{\mnras}} \textbf{\bibinfo{volume}{179}},
  \bibinfo{pages}{433--456} (\bibinfo{year}{1977}).

\bibitem{blandford1982}
\bibinfo{author}{{Blandford}, R.~D.} \& \bibinfo{author}{{Payne}, D.~G.}
\newblock \bibinfo{title}{{Hydromagnetic flows from accretion disks and the
  production of radio jets.}}
\newblock \emph{\bibinfo{journal}{\mnras}} \textbf{\bibinfo{volume}{199}},
  \bibinfo{pages}{883--903} (\bibinfo{year}{1982}).

\bibitem{hada2024}
\bibinfo{author}{{Hada}, K.}, \bibinfo{author}{{Asada}, K.},
  \bibinfo{author}{{Nakamura}, M.} \& \bibinfo{author}{{Kino}, M.}
\newblock \bibinfo{title}{{M 87: a cosmic laboratory for deciphering black hole
  accretion and jet formation}}.
\newblock \emph{\bibinfo{journal}{\aapr}} \textbf{\bibinfo{volume}{32}},
  \bibinfo{pages}{5} (\bibinfo{year}{2024}).

\bibitem{Lu:2023bbn}
\bibinfo{author}{Lu, R.-S.} \emph{et~al.}
\newblock \bibinfo{title}{{A ring-like accretion structure in M87 connecting
  its black hole and jet}}.
\newblock \emph{\bibinfo{journal}{Nature}} \textbf{\bibinfo{volume}{616}},
  \bibinfo{pages}{686--690} (\bibinfo{year}{2023}).

\bibitem{cui2023}
\bibinfo{author}{{Cui}, Y.} \emph{et~al.}
\newblock \bibinfo{title}{{Precessing jet nozzle connecting to a spinning black
  hole in M87}}.
\newblock \emph{\bibinfo{journal}{\nat}} \textbf{\bibinfo{volume}{621}},
  \bibinfo{pages}{711--715} (\bibinfo{year}{2023}).

\bibitem{lense1918}
\bibinfo{author}{{Lense}, J.} \& \bibinfo{author}{{Thirring}, H.}
\newblock \bibinfo{title}{{{\"U}ber den Einflu{\ss} der Eigenrotation der
  Zentralk{\"o}rper auf die Bewegung der Planeten und Monde nach der
  Einsteinschen Gravitationstheorie}}.
\newblock \emph{\bibinfo{journal}{Physikalische Zeitschrift}}
  \textbf{\bibinfo{volume}{19}}, \bibinfo{pages}{156} (\bibinfo{year}{1918}).

\bibitem{fragile2005}
\bibinfo{author}{{Fragile}, P.~C.} \& \bibinfo{author}{{Anninos}, P.}
\newblock \bibinfo{title}{{Hydrodynamic Simulations of Tilted Thick-Disk
  Accretion onto a Kerr Black Hole}}.
\newblock \emph{\bibinfo{journal}{\apj}} \textbf{\bibinfo{volume}{623}},
  \bibinfo{pages}{347--361} (\bibinfo{year}{2005}).

\bibitem{fragile2007}
\bibinfo{author}{{Fragile}, P.~C.}, \bibinfo{author}{{Blaes}, O.~M.},
  \bibinfo{author}{{Anninos}, P.} \& \bibinfo{author}{{Salmonson}, J.~D.}
\newblock \bibinfo{title}{{Global General Relativistic Magnetohydrodynamic
  Simulation of a Tilted Black Hole Accretion Disk}}.
\newblock \emph{\bibinfo{journal}{\apj}} \textbf{\bibinfo{volume}{668}},
  \bibinfo{pages}{417--429} (\bibinfo{year}{2007}).

\bibitem{mcKinney2013}
\bibinfo{author}{{McKinney}, J.~C.}, \bibinfo{author}{{Tchekhovskoy}, A.} \&
  \bibinfo{author}{{Blandford}, R.~D.}
\newblock \bibinfo{title}{{Alignment of Magnetized Accretion Disks and
  Relativistic Jets with Spinning Black Holes}}.
\newblock \emph{\bibinfo{journal}{Science}} \textbf{\bibinfo{volume}{339}},
  \bibinfo{pages}{49} (\bibinfo{year}{2013}).

\bibitem{kato1990}
\bibinfo{author}{{Kato}, S.}
\newblock \bibinfo{title}{{Trapped One-Armed Corrugation Waves and QPO's}}.
\newblock \emph{\bibinfo{journal}{\pasj}} \textbf{\bibinfo{volume}{42}},
  \bibinfo{pages}{99--113} (\bibinfo{year}{1990}).

\bibitem{Franchini:2015wna}
\bibinfo{author}{Franchini, A.}, \bibinfo{author}{Lodato, G.} \&
  \bibinfo{author}{Facchini, S.}
\newblock \bibinfo{title}{{Lense\textendash{}Thirring precession around
  supermassive black holes during tidal disruption events}}.
\newblock \emph{\bibinfo{journal}{Mon. Not. Roy. Astron. Soc.}}
  \textbf{\bibinfo{volume}{455}}, \bibinfo{pages}{1946--1956}
  (\bibinfo{year}{2016}).

\bibitem{bardeen1972}
\bibinfo{author}{{Bardeen}, J.~M.}, \bibinfo{author}{{Press}, W.~H.} \&
  \bibinfo{author}{{Teukolsky}, S.~A.}
\newblock \bibinfo{title}{{Rotating Black Holes: Locally Nonrotating Frames,
  Energy Extraction, and Scalar Synchrotron Radiation}}.
\newblock \emph{\bibinfo{journal}{\apj}} \textbf{\bibinfo{volume}{178}},
  \bibinfo{pages}{347--370} (\bibinfo{year}{1972}).

\bibitem{Wei:2024cti}
\bibinfo{author}{{Wei}, S.-W.}, \bibinfo{author}{{Zou}, Y.-C.},
  \bibinfo{author}{{Zhang}, Y.-P.} \& \bibinfo{author}{{Liu}, Y.-X.}
\newblock \bibinfo{title}{{Constraining black hole parameters with the
  precessing jet nozzle of M87*}}.
\newblock \emph{\bibinfo{journal}{\prd}} \textbf{\bibinfo{volume}{110}},
  \bibinfo{pages}{064006} (\bibinfo{year}{2024}).

\bibitem{Liu:2002wu}
\bibinfo{author}{Liu, S.-M.} \& \bibinfo{author}{Melia, F.}
\newblock \bibinfo{title}{{Spin - induced disk precession in the supermassive
  black hole at the Galactic Center}}.
\newblock \emph{\bibinfo{journal}{Astrophys. J. Lett.}}
  \textbf{\bibinfo{volume}{573}}, \bibinfo{pages}{L23} (\bibinfo{year}{2002}).

\bibitem{fragile2009}
\bibinfo{author}{{Fragile}, P.~C.}
\newblock \bibinfo{title}{{Effective Inner Radius of Tilted Black Hole
  Accretion Disks}}.
\newblock \emph{\bibinfo{journal}{\apjl}} \textbf{\bibinfo{volume}{706}},
  \bibinfo{pages}{L246--L250} (\bibinfo{year}{2009}).

\bibitem{sheperd2023}
\bibinfo{author}{{Doeleman}, S.~S.} \emph{et~al.}
\newblock \bibinfo{title}{{Reference Array and Design Consideration for the
  Next-Generation Event Horizon Telescope}}.
\newblock \emph{\bibinfo{journal}{Galaxies}} \textbf{\bibinfo{volume}{11}},
  \bibinfo{pages}{107} (\bibinfo{year}{2023}).

\bibitem{Palumbo_2020}
\bibinfo{author}{{Palumbo}, D. C.~M.}, \bibinfo{author}{{Wong}, G.~N.} \&
  \bibinfo{author}{{Prather}, B.~S.}
\newblock \bibinfo{title}{{Discriminating Accretion States via Rotational
  Symmetry in Simulated Polarimetric Images of M87}}.
\newblock \emph{\bibinfo{journal}{\apj}} \textbf{\bibinfo{volume}{894}},
  \bibinfo{pages}{156} (\bibinfo{year}{2020}).

\bibitem{Chael_2023}
\bibinfo{author}{{Chael}, A.}, \bibinfo{author}{{Lupsasca}, A.},
  \bibinfo{author}{{Wong}, G.~N.} \& \bibinfo{author}{{Quataert}, E.}
\newblock \bibinfo{title}{{Black Hole Polarimetry I. A Signature of
  Electromagnetic Energy Extraction}}.
\newblock \emph{\bibinfo{journal}{\apj}} \textbf{\bibinfo{volume}{958}},
  \bibinfo{pages}{65} (\bibinfo{year}{2023}).

\bibitem{Ricarte:2022kft}
\bibinfo{author}{Ricarte, A.}, \bibinfo{author}{Tiede, P.},
  \bibinfo{author}{Emami, R.}, \bibinfo{author}{Tamar, A.} \&
  \bibinfo{author}{Natarajan, P.}
\newblock \bibinfo{title}{{The ngEHT\textquoteright{}s Role in Measuring
  Supermassive Black Hole Spins}}.
\newblock \emph{\bibinfo{journal}{Galaxies}} \textbf{\bibinfo{volume}{11}},
  \bibinfo{pages}{6} (\bibinfo{year}{2023}).

\bibitem{asada2012}
\bibinfo{author}{{Asada}, K.} \& \bibinfo{author}{{Nakamura}, M.}
\newblock \bibinfo{title}{{The Structure of the M87 Jet: A Transition from
  Parabolic to Conical Streamlines}}.
\newblock \emph{\bibinfo{journal}{\apjl}} \textbf{\bibinfo{volume}{745}},
  \bibinfo{pages}{L28} (\bibinfo{year}{2012}).

\bibitem{nakamura2018}
\bibinfo{author}{{Nakamura}, M.} \emph{et~al.}
\newblock \bibinfo{title}{{Parabolic Jets from the Spinning Black Hole in
  M87}}.
\newblock \emph{\bibinfo{journal}{\apj}} \textbf{\bibinfo{volume}{868}},
  \bibinfo{pages}{146} (\bibinfo{year}{2018}).

\bibitem{mwl2021}
\bibinfo{author}{{EHT MWL Science Working Group}} \emph{et~al.}
\newblock \bibinfo{title}{{Broadband Multi-wavelength Properties of M87 during
  the 2017 Event Horizon Telescope Campaign}}.
\newblock \emph{\bibinfo{journal}{\apjl}} \textbf{\bibinfo{volume}{911}},
  \bibinfo{pages}{L11} (\bibinfo{year}{2021}).

\bibitem{nikonov2023}
\bibinfo{author}{{Nikonov}, A.~S.}, \bibinfo{author}{{Kovalev}, Y.~Y.},
  \bibinfo{author}{{Kravchenko}, E.~V.}, \bibinfo{author}{{Pashchenko}, I.~N.}
  \& \bibinfo{author}{{Lobanov}, A.~P.}
\newblock \bibinfo{title}{{Properties of the jet in M87 revealed by its helical
  structure imaged with the VLBA at 8 and 15 GHz}}.
\newblock \emph{\bibinfo{journal}{\mnras}} \textbf{\bibinfo{volume}{526}},
  \bibinfo{pages}{5949--5963} (\bibinfo{year}{2023}).

\bibitem{MWL2024}
\bibinfo{author}{{EHT MWL Science Working Group}} \emph{et~al.}
\newblock \bibinfo{title}{{Broadband Multi-wavelength Properties of M87 during
  the 2018 EHT Campaign including a Very High Energy Flaring Episode}}.
\newblock \emph{\bibinfo{journal}{\aap}} \bibinfo{pages}{692}
  (\bibinfo{year}{2024}).

\bibitem{liska2018}
\bibinfo{author}{{Liska}, M.} \emph{et~al.}
\newblock \bibinfo{title}{{Formation of precessing jets by tilted black hole
  discs in 3D general relativistic MHD simulations}}.
\newblock \emph{\bibinfo{journal}{\mnras}} \textbf{\bibinfo{volume}{474}},
  \bibinfo{pages}{L81--L85} (\bibinfo{year}{2018}).

\bibitem{Rohoza:2023egi}
\bibinfo{author}{Rohoza, V.} \emph{et~al.}
\newblock \bibinfo{title}{{How to Turn Jets into Cylinders near Supermassive
  Black Holes in 3D General Relativistic Magnetohydrodynamic Simulations}}.
\newblock \emph{\bibinfo{journal}{Astrophys. J. Lett.}}
  \textbf{\bibinfo{volume}{963}}, \bibinfo{pages}{L29} (\bibinfo{year}{2024}).

\bibitem{hada2016gmva}
\bibinfo{author}{{Hada}, K.} \emph{et~al.}
\newblock \bibinfo{title}{{High-sensitivity 86 GHz (3.5 mm) VLBI Observations
  of M87: Deep Imaging of the Jet Base at a Resolution of 10 Schwarzschild
  Radii}}.
\newblock \emph{\bibinfo{journal}{\apj}} \textbf{\bibinfo{volume}{817}},
  \bibinfo{pages}{131} (\bibinfo{year}{2016}).

\bibitem{kim2018gmva}
\bibinfo{author}{{Kim}, J.~Y.} \emph{et~al.}
\newblock \bibinfo{title}{{The limb-brightened jet of M87 down to the 7
  Schwarzschild radii scale}}.
\newblock \emph{\bibinfo{journal}{\aap}} \textbf{\bibinfo{volume}{616}},
  \bibinfo{pages}{A188} (\bibinfo{year}{2018}).

\bibitem{Porth_2019}
\bibinfo{author}{{Porth}, O.} \emph{et~al.}
\newblock \bibinfo{title}{{The Event Horizon General Relativistic
  Magnetohydrodynamic Code Comparison Project}}.
\newblock \emph{\bibinfo{journal}{\apjs}} \textbf{\bibinfo{volume}{243}},
  \bibinfo{pages}{26} (\bibinfo{year}{2019}).

\bibitem{chatterjee2022}
\bibinfo{author}{{Chatterjee}, K.} \& \bibinfo{author}{{Narayan}, R.}
\newblock \bibinfo{title}{{Flux Eruption Events Drive Angular Momentum
  Transport in Magnetically Arrested Accretion Flows}}.
\newblock \emph{\bibinfo{journal}{\apj}} \textbf{\bibinfo{volume}{941}},
  \bibinfo{pages}{30} (\bibinfo{year}{2022}).

\bibitem{Tchekhovskoy2015}
\bibinfo{author}{Tchekhovskoy, A.}
\newblock \emph{\bibinfo{title}{Launching of Active Galactic Nuclei Jets}},
  \bibinfo{pages}{45--82} (\bibinfo{publisher}{Springer International
  Publishing}, \bibinfo{address}{Cham}, \bibinfo{year}{2015}).

\bibitem{Gupta:2024shd}
\bibinfo{author}{{Gupta}, S.} \& \bibinfo{author}{{Dexter}, J.}
\newblock \bibinfo{title}{{Shock-induced Partial Alignment in Geometrically
  Thick Tilted Accretion Disks Around Black Holes}}.
\newblock \emph{\bibinfo{journal}{\apj}} \textbf{\bibinfo{volume}{974}},
  \bibinfo{pages}{209} (\bibinfo{year}{2024}).

\bibitem{eht92023}
\bibinfo{author}{{Event Horizon Telescope Collaboration}}.
\newblock \bibinfo{title}{{First M87 Event Horizon Telescope Results. IX.
  Detection of Near-horizon Circular Polarization}}.
\newblock \emph{\bibinfo{journal}{\apjl}} \textbf{\bibinfo{volume}{957}},
  \bibinfo{pages}{L20} (\bibinfo{year}{2023}).

\bibitem{fragile2024}
\bibinfo{author}{{Fragile}, P.~C.} \& \bibinfo{author}{{Liska}, M.}
\newblock \emph{\bibinfo{title}{{Tilted Accretion Disks}}},
  \bibinfo{pages}{361--387} (\bibinfo{publisher}{Springer Nature Singapore},
  \bibinfo{address}{Singapore}, \bibinfo{year}{2025}).

\bibitem{janssen2019}
\bibinfo{author}{{Janssen}, M.} \emph{et~al.}
\newblock \bibinfo{title}{{rPICARD: A CASA-based calibration pipeline for VLBI
  data. Calibration and imaging of 7 mm VLBA observations of the AGN jet in M
  87}}.
\newblock \emph{\bibinfo{journal}{\aap}} \textbf{\bibinfo{volume}{626}},
  \bibinfo{pages}{A75} (\bibinfo{year}{2019}).

\bibitem{chael2019}
\bibinfo{author}{{Chael}, A.~A.} \emph{et~al.}
\newblock \bibinfo{title}{{ehtim: Imaging, analysis, and simulation software
  for radio interferometry}}.
\newblock \bibinfo{howpublished}{Astrophysics Source Code Library, record
  ascl:1904.004} (\bibinfo{year}{2019}).

\bibitem{tiede2022}
\bibinfo{author}{{Tiede}, P.}
\newblock \bibinfo{title}{{Comrade: Composable Modeling of Radio Emission}}.
\newblock \emph{\bibinfo{journal}{The Journal of Open Source Software}}
  \textbf{\bibinfo{volume}{7}}, \bibinfo{pages}{4457} (\bibinfo{year}{2022}).

\bibitem{tazaki2023}
\bibinfo{author}{{Tazaki}, F.} \emph{et~al.}
\newblock \bibinfo{title}{{Super-Resolved Image of M87 Observed with East Asian
  VLBI Network}}.
\newblock \emph{\bibinfo{journal}{Galaxies}} \textbf{\bibinfo{volume}{11}},
  \bibinfo{pages}{39} (\bibinfo{year}{2023}).

\bibitem{kim2024a}
\bibinfo{author}{{Kim}, J.-S.} \emph{et~al.}
\newblock \bibinfo{title}{{Bayesian self-calibration and imaging in very long
  baseline interferometry}}.
\newblock \emph{\bibinfo{journal}{\aap}} \textbf{\bibinfo{volume}{690}},
  \bibinfo{pages}{A129} (\bibinfo{year}{2024}).

\bibitem{Lubow+2002}
\bibinfo{author}{Lubow, S.~H.}, \bibinfo{author}{Ogilvie, G.~I.} \&
  \bibinfo{author}{Pringle, J.~E.}
\newblock \bibinfo{title}{The evolution of a warped disc around a kerr black
  hole}.
\newblock \emph{\bibinfo{journal}{\mnras}} \textbf{\bibinfo{volume}{337}},
  \bibinfo{pages}{706--712} (\bibinfo{year}{2002}).

\bibitem{lodato2013}
\bibinfo{author}{{Lodato}, G.} \& \bibinfo{author}{{Facchini}, S.}
\newblock \bibinfo{title}{{Wave-like warp propagation in circumbinary discs -
  II. Application to KH 15D}}.
\newblock \emph{\bibinfo{journal}{\mnras}} \textbf{\bibinfo{volume}{433}},
  \bibinfo{pages}{2157--2164} (\bibinfo{year}{2013}).

\bibitem{nixon2012}
\bibinfo{author}{{Nixon}, C.~J.} \& \bibinfo{author}{{King}, A.~R.}
\newblock \bibinfo{title}{{Broken discs: warp propagation in accretion discs}}.
\newblock \emph{\bibinfo{journal}{\mnras}} \textbf{\bibinfo{volume}{421}},
  \bibinfo{pages}{1201--1208} (\bibinfo{year}{2012}).

\bibitem{wielgus2020}
\bibinfo{author}{{Wielgus}, M.} \emph{et~al.}
\newblock \bibinfo{title}{{Monitoring the Morphology of M87* in 2009-2017 with
  the Event Horizon Telescope}}.
\newblock \emph{\bibinfo{journal}{\apj}} \textbf{\bibinfo{volume}{901}},
  \bibinfo{pages}{67} (\bibinfo{year}{2020}).

\bibitem{Geroch:1970cd}
\bibinfo{author}{Geroch, R.~P.}
\newblock \bibinfo{title}{{Multipole moments. II. Curved space}}.
\newblock \emph{\bibinfo{journal}{J. Math. Phys.}}
  \textbf{\bibinfo{volume}{11}}, \bibinfo{pages}{2580--2588}
  (\bibinfo{year}{1970}).

\bibitem{Glampedakis:2005cf}
\bibinfo{author}{Glampedakis, K.} \& \bibinfo{author}{Babak, S.}
\newblock \bibinfo{title}{{Mapping spacetimes with LISA: Inspiral of a
  test-body in a `quasi-Kerr' field}}.
\newblock \emph{\bibinfo{journal}{Class. Quant. Grav.}}
  \textbf{\bibinfo{volume}{23}}, \bibinfo{pages}{4167--4188}
  (\bibinfo{year}{2006}).

\bibitem{Yagi:2016jml}
\bibinfo{author}{Yagi, K.} \& \bibinfo{author}{Stein, L.~C.}
\newblock \bibinfo{title}{{Black Hole Based Tests of General Relativity}}.
\newblock \emph{\bibinfo{journal}{Class. Quant. Grav.}}
  \textbf{\bibinfo{volume}{33}}, \bibinfo{pages}{054001}
  (\bibinfo{year}{2016}).

\bibitem{Alexander:2009tp}
\bibinfo{author}{Alexander, S.} \& \bibinfo{author}{Yunes, N.}
\newblock \bibinfo{title}{{Chern-Simons Modified General Relativity}}.
\newblock \emph{\bibinfo{journal}{Phys. Rept.}} \textbf{\bibinfo{volume}{480}},
  \bibinfo{pages}{1--55} (\bibinfo{year}{2009}).

\bibitem{mertens2016}
\bibinfo{author}{{Mertens}, F.}, \bibinfo{author}{{Lobanov}, A.~P.},
  \bibinfo{author}{{Walker}, R.~C.} \& \bibinfo{author}{{Hardee}, P.~E.}
\newblock \bibinfo{title}{{Kinematics of the jet in M 87 on scales of 100-1000
  Schwarzschild radii}}.
\newblock \emph{\bibinfo{journal}{\aap}} \textbf{\bibinfo{volume}{595}},
  \bibinfo{pages}{A54} (\bibinfo{year}{2016}).

\bibitem{walker2018}
\bibinfo{author}{{Walker}, R.~C.}, \bibinfo{author}{{Hardee}, P.~E.},
  \bibinfo{author}{{Davies}, F.~B.}, \bibinfo{author}{{Ly}, C.} \&
  \bibinfo{author}{{Junor}, W.}
\newblock \bibinfo{title}{{The Structure and Dynamics of the Subparsec Jet in
  M87 Based on 50 VLBA Observations over 17 Years at 43 GHz}}.
\newblock \emph{\bibinfo{journal}{\apj}} \textbf{\bibinfo{volume}{855}},
  \bibinfo{pages}{128} (\bibinfo{year}{2018}).

\bibitem{chatterjee2019}
\bibinfo{author}{{Chatterjee}, K.}, \bibinfo{author}{{Liska}, M.},
  \bibinfo{author}{{Tchekhovskoy}, A.} \& \bibinfo{author}{{Markoff}, S.~B.}
\newblock \bibinfo{title}{{Accelerating AGN jets to parsec scales using general
  relativistic MHD simulations}}.
\newblock \emph{\bibinfo{journal}{\mnras}} \textbf{\bibinfo{volume}{490}},
  \bibinfo{pages}{2200--2218} (\bibinfo{year}{2019}).

\bibitem{pu2016}
\bibinfo{author}{{Pu}, H.-Y.}, \bibinfo{author}{{Yun}, K.},
  \bibinfo{author}{{Younsi}, Z.} \& \bibinfo{author}{{Yoon}, S.-J.}
\newblock \bibinfo{title}{{Odyssey: A Public GPU-based Code for General
  Relativistic Radiative Transfer in Kerr Spacetime}}.
\newblock \emph{\bibinfo{journal}{\apj}} \textbf{\bibinfo{volume}{820}},
  \bibinfo{pages}{105} (\bibinfo{year}{2016}).

\bibitem{liska2019}
\bibinfo{author}{{Liska}, M.}, \bibinfo{author}{{Tchekhovskoy}, A.},
  \bibinfo{author}{{Ingram}, A.} \& \bibinfo{author}{{van der Klis}, M.}
\newblock \bibinfo{title}{{Bardeen-Petterson alignment, jets, and magnetic
  truncation in GRMHD simulations of tilted thin accretion discs}}.
\newblock \emph{\bibinfo{journal}{\mnras}} \textbf{\bibinfo{volume}{487}},
  \bibinfo{pages}{550--561} (\bibinfo{year}{2019}).

\bibitem{narayan2003}
\bibinfo{author}{{Narayan}, R.}, \bibinfo{author}{{Igumenshchev}, I.~V.} \&
  \bibinfo{author}{{Abramowicz}, M.~A.}
\newblock \bibinfo{title}{{Magnetically Arrested Disk: an Energetically
  Efficient Accretion Flow}}.
\newblock \emph{\bibinfo{journal}{\pasj}} \textbf{\bibinfo{volume}{55}},
  \bibinfo{pages}{L69--L72} (\bibinfo{year}{2003}).

\bibitem{tchek2011}
\bibinfo{author}{{Tchekhovskoy}, A.}, \bibinfo{author}{{Narayan}, R.} \&
  \bibinfo{author}{{McKinney}, J.~C.}
\newblock \bibinfo{title}{{Efficient generation of jets from magnetically
  arrested accretion on a rapidly spinning black hole}}.
\newblock \emph{\bibinfo{journal}{\mnras}} \textbf{\bibinfo{volume}{418}},
  \bibinfo{pages}{L79--L83} (\bibinfo{year}{2011}).

\bibitem{mcKinney2012}
\bibinfo{author}{{McKinney}, J.~C.}, \bibinfo{author}{{Tchekhovskoy}, A.} \&
  \bibinfo{author}{{Bland ford}, R.~D.}
\newblock \bibinfo{title}{{General relativistic magnetohydrodynamic simulations
  of magnetically choked accretion flows around black holes}}.
\newblock \emph{\bibinfo{journal}{\mnras}} \textbf{\bibinfo{volume}{423}},
  \bibinfo{pages}{3083--3117} (\bibinfo{year}{2012}).

\bibitem{narayan2022}
\bibinfo{author}{{Narayan}, R.}, \bibinfo{author}{{Chael}, A.},
  \bibinfo{author}{{Chatterjee}, K.}, \bibinfo{author}{{Ricarte}, A.} \&
  \bibinfo{author}{{Curd}, B.}
\newblock \bibinfo{title}{{Jets in magnetically arrested hot accretion flows:
  geometry, power, and black hole spin-down}}.
\newblock \emph{\bibinfo{journal}{\mnras}} \textbf{\bibinfo{volume}{511}},
  \bibinfo{pages}{3795--3813} (\bibinfo{year}{2022}).

\bibitem{DeVilliers2003}
\bibinfo{author}{{De Villiers}, J.-P.}, \bibinfo{author}{{Hawley}, J.~F.} \&
  \bibinfo{author}{{Krolik}, J.~H.}
\newblock \bibinfo{title}{{Magnetically Driven Accretion Flows in the Kerr
  Metric. I. Models and Overall Structure}}.
\newblock \emph{\bibinfo{journal}{\apj}} \textbf{\bibinfo{volume}{599}},
  \bibinfo{pages}{1238--1253} (\bibinfo{year}{2003}).

\bibitem{gammie2003}
\bibinfo{author}{{Gammie}, C.~F.}, \bibinfo{author}{{McKinney}, J.~C.} \&
  \bibinfo{author}{{T{\'o}th}, G.}
\newblock \bibinfo{title}{{HARM: A Numerical Scheme for General Relativistic
  Magnetohydrodynamics}}.
\newblock \emph{\bibinfo{journal}{\apj}} \textbf{\bibinfo{volume}{589}},
  \bibinfo{pages}{444--457} (\bibinfo{year}{2003}).

\bibitem{narayan2012}
\bibinfo{author}{{Narayan}, R.}, \bibinfo{author}{{S{\"A} dowski}, A.},
  \bibinfo{author}{{Penna}, R.~F.} \& \bibinfo{author}{{Kulkarni}, A.~K.}
\newblock \bibinfo{title}{{GRMHD simulations of magnetized advection-dominated
  accretion on a non-spinning black hole: role of outflows}}.
\newblock \emph{\bibinfo{journal}{\mnras}} \textbf{\bibinfo{volume}{426}},
  \bibinfo{pages}{3241--3259} (\bibinfo{year}{2012}).

\bibitem{sadowski2013}
\bibinfo{author}{{S{\k{a}}dowski}, A.}, \bibinfo{author}{{Narayan}, R.},
  \bibinfo{author}{{Penna}, R.} \& \bibinfo{author}{{Zhu}, Y.}
\newblock \bibinfo{title}{{Energy, momentum and mass outflows and feedback from
  thick accretion discs around rotating black holes}}.
\newblock \emph{\bibinfo{journal}{\mnras}} \textbf{\bibinfo{volume}{436}},
  \bibinfo{pages}{3856--3874} (\bibinfo{year}{2013}).

\bibitem{aktar2024}
\bibinfo{author}{{Aktar}, R.}, \bibinfo{author}{{Pan}, K.-C.} \&
  \bibinfo{author}{{Okuda}, T.}
\newblock \bibinfo{title}{{Radiation RMHD Accretion Flows around Spinning AGNs:
  A Comparative Study of MAD and SANE State}}.
\newblock \emph{\bibinfo{journal}{\apj}} \textbf{\bibinfo{volume}{972}},
  \bibinfo{pages}{18} (\bibinfo{year}{2024}).

\bibitem{yao2021}
\bibinfo{author}{{Yao}, P.~Z.}, \bibinfo{author}{{Dexter}, J.},
  \bibinfo{author}{{Chen}, A.~Y.}, \bibinfo{author}{{Ryan}, B.~R.} \&
  \bibinfo{author}{{Wong}, G.~N.}
\newblock \bibinfo{title}{{Radiation GRMHD simulations of M87: funnel
  properties and prospects for gap acceleration}}.
\newblock \emph{\bibinfo{journal}{\mnras}} \textbf{\bibinfo{volume}{507}},
  \bibinfo{pages}{4864--4878} (\bibinfo{year}{2021}).

\bibitem{Tsunetoe:2022ktx}
\bibinfo{author}{Tsunetoe, Y.} \emph{et~al.}
\newblock \bibinfo{title}{{Investigating the Disk\textendash{}Jet Structure in
  M87 through Flux Separation in the Linear and Circular Polarization Images}}.
\newblock \emph{\bibinfo{journal}{Astrophys. J.}}
  \textbf{\bibinfo{volume}{931}}, \bibinfo{pages}{25} (\bibinfo{year}{2022}).

\bibitem{white2019}
\bibinfo{author}{{White}, C.~J.}, \bibinfo{author}{{Quataert}, E.} \&
  \bibinfo{author}{{Blaes}, O.}
\newblock \bibinfo{title}{{Tilted Disks around Black Holes: A Numerical
  Parameter Survey for Spin and Inclination Angle}}.
\newblock \emph{\bibinfo{journal}{\apj}} \textbf{\bibinfo{volume}{878}},
  \bibinfo{pages}{51} (\bibinfo{year}{2019}).

\bibitem{ressler2021}
\bibinfo{author}{{Ressler}, S.~M.}, \bibinfo{author}{{Quataert}, E.},
  \bibinfo{author}{{White}, C.~J.} \& \bibinfo{author}{{Blaes}, O.}
\newblock \bibinfo{title}{{Magnetically modified spherical accretion in GRMHD:
  reconnection-driven convection and jet propagation}}.
\newblock \emph{\bibinfo{journal}{\mnras}} \textbf{\bibinfo{volume}{504}},
  \bibinfo{pages}{6076--6095} (\bibinfo{year}{2021}).

\bibitem{liska2021}
\bibinfo{author}{{Liska}, M.} \emph{et~al.}
\newblock \bibinfo{title}{{Disc tearing and Bardeen-Petterson alignment in
  GRMHD simulations of highly tilted thin accretion discs}}.
\newblock \emph{\bibinfo{journal}{\mnras}} \textbf{\bibinfo{volume}{507}},
  \bibinfo{pages}{983--990} (\bibinfo{year}{2021}).

\bibitem{Chatterjee_etal:2023}
\bibinfo{author}{{Chatterjee}, K.}, \bibinfo{author}{{Liska}, M.},
  \bibinfo{author}{{Tchekhovskoy}, A.} \& \bibinfo{author}{{Markoff}, S.}
\newblock \bibinfo{title}{{Misaligned magnetized accretion flows onto spinning
  black holes: magneto-spin alignment, outflow power and intermittent jets}}.
\newblock \emph{\bibinfo{journal}{arXiv e-prints}}
  \bibinfo{pages}{arXiv:2311.00432} (\bibinfo{year}{2023}).

\bibitem{ressler2023}
\bibinfo{author}{{Ressler}, S.~M.}, \bibinfo{author}{{White}, C.~J.} \&
  \bibinfo{author}{{Quataert}, E.}
\newblock \bibinfo{title}{{Wind-fed GRMHD simulations of Sagittarius A*: tilt
  and alignment of jets and accretion discs, electron thermodynamics, and
  multiscale modelling of the rotation measure}}.
\newblock \emph{\bibinfo{journal}{\mnras}} \textbf{\bibinfo{volume}{521}},
  \bibinfo{pages}{4277--4298} (\bibinfo{year}{2023}).

\bibitem{ressler2024MAD}
\bibinfo{author}{{Ressler}, S.~M.}, \bibinfo{author}{{Combi}, L.},
  \bibinfo{author}{{Li}, X.}, \bibinfo{author}{{Ripperda}, B.} \&
  \bibinfo{author}{{Yang}, H.}
\newblock \bibinfo{title}{{Black Hole{\textendash}Disk Interactions in
  Magnetically Arrested Active Galactic Nuclei: General Relativistic
  Magnetohydrodynamic Simulations Using a Time-dependent, Binary Metric}}.
\newblock \emph{\bibinfo{journal}{\apj}} \textbf{\bibinfo{volume}{967}},
  \bibinfo{pages}{70} (\bibinfo{year}{2024}).

\bibitem{bollimpalli2023}
\bibinfo{author}{{Bollimpalli}, D.~A.}, \bibinfo{author}{{Fragile}, P.~C.} \&
  \bibinfo{author}{{Klu{\'z}niak}, W.}
\newblock \bibinfo{title}{{Effect of geometrically thin discs on precessing,
  thick flows: relevance to type-C QPOs}}.
\newblock \emph{\bibinfo{journal}{\mnras}} \textbf{\bibinfo{volume}{520}},
  \bibinfo{pages}{L79--L84} (\bibinfo{year}{2023}).

\bibitem{bollimpalli2024}
\bibinfo{author}{{Bollimpalli}, D.~A.}, \bibinfo{author}{{Fragile}, P.~C.},
  \bibinfo{author}{{Dewberry}, J.~W.} \& \bibinfo{author}{{Klu{\'z}niak}, W.}
\newblock \bibinfo{title}{{Truncated, tilted discs as a possible source of
  Quasi-Periodic Oscillations}}.
\newblock \emph{\bibinfo{journal}{\mnras}} \textbf{\bibinfo{volume}{528}},
  \bibinfo{pages}{1142--1157} (\bibinfo{year}{2024}).

\bibitem{dogan2018}
\bibinfo{author}{{Do{\v{g}}an}, S.}, \bibinfo{author}{{Nixon}, C.~J.},
  \bibinfo{author}{{King}, A.~R.} \& \bibinfo{author}{{Pringle}, J.~E.}
\newblock \bibinfo{title}{{Instability of warped discs}}.
\newblock \emph{\bibinfo{journal}{\mnras}} \textbf{\bibinfo{volume}{476}},
  \bibinfo{pages}{1519--1531} (\bibinfo{year}{2018}).

\bibitem{Fragile:2023qpz}
\bibinfo{author}{Fragile, P.~C.}, \bibinfo{author}{Chatterjee, K.},
  \bibinfo{author}{Ingram, A.} \& \bibinfo{author}{Middleton, M.}
\newblock \bibinfo{title}{{The luminous, hard state can\textquoteright{}t be
  MAD}}.
\newblock \emph{\bibinfo{journal}{Mon. Not. Roy. Astron. Soc.}}
  \textbf{\bibinfo{volume}{525}}, \bibinfo{pages}{L82--L86}
  (\bibinfo{year}{2023}).

\bibitem{begelman1980}
\bibinfo{author}{{Begelman}, M.~C.}, \bibinfo{author}{{Blandford}, R.~D.} \&
  \bibinfo{author}{{Rees}, M.~J.}
\newblock \bibinfo{title}{{Massive black hole binaries in active galactic
  nuclei}}.
\newblock \emph{\bibinfo{journal}{\nat}} \textbf{\bibinfo{volume}{287}},
  \bibinfo{pages}{307--309} (\bibinfo{year}{1980}).

\bibitem{valtonen2008}
\bibinfo{author}{{Valtonen}, M.~J.} \emph{et~al.}
\newblock \bibinfo{title}{{A massive binary black-hole system in OJ287 and a
  test of general relativity}}.
\newblock \emph{\bibinfo{journal}{\nat}} \textbf{\bibinfo{volume}{452}},
  \bibinfo{pages}{851--853} (\bibinfo{year}{2008}).

\bibitem{ressler2025BBH}
\bibinfo{author}{{Ressler}, S.~M.}, \bibinfo{author}{{Combi}, L.},
  \bibinfo{author}{{Ripperda}, B.} \& \bibinfo{author}{{Most}, E.~R.}
\newblock \bibinfo{title}{{Dual Jet Interaction, Magnetically Arrested Flows,
  and Flares in Accreting Binary Black Holes}}.
\newblock \emph{\bibinfo{journal}{\apjl}} \textbf{\bibinfo{volume}{979}},
  \bibinfo{pages}{L24} (\bibinfo{year}{2025}).

\bibitem{Shen:2024gwb}
\bibinfo{author}{{Shen}, Y.} \& \bibinfo{author}{{Chen}, B.}
\newblock \bibinfo{title}{{Precession and split of tilted, geometrically thin
  accretion disk: an analytical study}}.
\newblock \emph{\bibinfo{journal}{\jcap}} \textbf{\bibinfo{volume}{2024}},
  \bibinfo{pages}{063} (\bibinfo{year}{2024}).

\bibitem{acciari2009}
\bibinfo{author}{{Acciari}, V.~A.} \emph{et~al.}
\newblock \bibinfo{title}{{Radio Imaging of the Very-High-Energy
  {\ensuremath{\gamma}}-Ray Emission Region in the Central Engine of a Radio
  Galaxy}}.
\newblock \emph{\bibinfo{journal}{Science}} \textbf{\bibinfo{volume}{325}},
  \bibinfo{pages}{444} (\bibinfo{year}{2009}).

\bibitem{mizuno2009}
\bibinfo{author}{{Mizuno}, Y.}, \bibinfo{author}{{Lyubarsky}, Y.},
  \bibinfo{author}{{Nishikawa}, K.-I.} \& \bibinfo{author}{{Hardee}, P.~E.}
\newblock \bibinfo{title}{{Three-Dimensional Relativistic Magnetohydrodynamic
  Simulations of Current-Driven Instability. I. Instability of a Static
  Column}}.
\newblock \emph{\bibinfo{journal}{\apj}} \textbf{\bibinfo{volume}{700}},
  \bibinfo{pages}{684--693} (\bibinfo{year}{2009}).

\bibitem{mizuno2012}
\bibinfo{author}{{Mizuno}, Y.}, \bibinfo{author}{{Lyubarsky}, Y.},
  \bibinfo{author}{{Nishikawa}, K.-I.} \& \bibinfo{author}{{Hardee}, P.~E.}
\newblock \bibinfo{title}{{Three-dimensional Relativistic Magnetohydrodynamic
  Simulations of Current-driven Instability. III. Rotating Relativistic Jets}}.
\newblock \emph{\bibinfo{journal}{\apj}} \textbf{\bibinfo{volume}{757}},
  \bibinfo{pages}{16} (\bibinfo{year}{2012}).

\bibitem{blandford1976}
\bibinfo{author}{{Blandford}, R.~D.} \& \bibinfo{author}{{Pringle}, J.~E.}
\newblock \bibinfo{title}{{Kelvin-Helmholtz instability of relativistic
  beams.}}
\newblock \emph{\bibinfo{journal}{\mnras}} \textbf{\bibinfo{volume}{176}},
  \bibinfo{pages}{443--454} (\bibinfo{year}{1976}).

\bibitem{mizuno2007}
\bibinfo{author}{{Mizuno}, Y.}, \bibinfo{author}{{Hardee}, P.} \&
  \bibinfo{author}{{Nishikawa}, K.-I.}
\newblock \bibinfo{title}{{Three-dimensional Relativistic Magnetohydrodynamic
  Simulations of Magnetized Spine-Sheath Relativistic Jets}}.
\newblock \emph{\bibinfo{journal}{\apj}} \textbf{\bibinfo{volume}{662}},
  \bibinfo{pages}{835--850} (\bibinfo{year}{2007}).

\bibitem{lobanov2003}
\bibinfo{author}{{Lobanov}, A.}, \bibinfo{author}{{Hardee}, P.} \&
  \bibinfo{author}{{Eilek}, J.}
\newblock \bibinfo{title}{{Internal structure and dynamics of the
  kiloparsec-scale jet in M87}}.
\newblock \emph{\bibinfo{journal}{\nar}} \textbf{\bibinfo{volume}{47}},
  \bibinfo{pages}{629--632} (\bibinfo{year}{2003}).

\bibitem{pasetto2021}
\bibinfo{author}{{Pasetto}, A.} \emph{et~al.}
\newblock \bibinfo{title}{{Reading M87's DNA: A Double Helix Revealing a
  Large-scale Helical Magnetic Field}}.
\newblock \emph{\bibinfo{journal}{\apjl}} \textbf{\bibinfo{volume}{923}},
  \bibinfo{pages}{L5} (\bibinfo{year}{2021}).

\bibitem{tsunetoe2024}
\bibinfo{author}{Tsunetoe, Y.}, \bibinfo{author}{Narayan, R.} \&
  \bibinfo{author}{Ricarte, A.}
\newblock \bibinfo{title}{Jet archaeology and forecasting: Image variability
  and magnetic field configuration}.
\newblock \emph{\bibinfo{journal}{\apj}} \textbf{\bibinfo{volume}{983}},
  \bibinfo{pages}{77} (\bibinfo{year}{2025}).

\bibitem{Lalakos:2023ean}
\bibinfo{author}{Lalakos, A.} \emph{et~al.}
\newblock \bibinfo{title}{{Jets with a Twist: The Emergence of FR0 Jets in a 3D
  GRMHD Simulation of Zero-angular-momentum Black Hole Accretion}}.
\newblock \emph{\bibinfo{journal}{Astrophys. J.}}
  \textbf{\bibinfo{volume}{964}}, \bibinfo{pages}{79} (\bibinfo{year}{2024}).

\bibitem{Giovannini2023}
\bibinfo{author}{{Giovannini}, G.} \emph{et~al.}
\newblock \bibinfo{title}{{The Past and Future of East Asia to Italy: Nearly
  Global VLBI}}.
\newblock \emph{\bibinfo{journal}{Galaxies}} \textbf{\bibinfo{volume}{11}},
  \bibinfo{pages}{49} (\bibinfo{year}{2023}).

\bibitem{Pushkarev2017}
\bibinfo{author}{{Pushkarev}, A.~B.}, \bibinfo{author}{{Kovalev}, Y.~Y.},
  \bibinfo{author}{{Lister}, M.~L.} \& \bibinfo{author}{{Savolainen}, T.}
\newblock \bibinfo{title}{{MOJAVE - XIV. Shapes and opening angles of AGN
  jets}}.
\newblock \emph{\bibinfo{journal}{\mnras}} \textbf{\bibinfo{volume}{468}},
  \bibinfo{pages}{4992--5003} (\bibinfo{year}{2017}).

\bibitem{akiyama2017}
\bibinfo{author}{{Akiyama}, K.} \emph{et~al.}
\newblock \bibinfo{title}{{Superresolution Full-polarimetric Imaging for Radio
  Interferometry with Sparse Modeling}}.
\newblock \emph{\bibinfo{journal}{\aj}} \textbf{\bibinfo{volume}{153}},
  \bibinfo{pages}{159} (\bibinfo{year}{2017}).

\bibitem{kim2024b}
\bibinfo{author}{{Kim}, J.-S.} \emph{et~al.}
\newblock \bibinfo{title}{{Imaging a ring-like structure and the extended jet
  of M87 at 86 GHz}}.
\newblock \emph{\bibinfo{journal}{\aap}} \textbf{\bibinfo{volume}{696}},
  \bibinfo{pages}{A169} (\bibinfo{year}{2025}).

\bibitem{lisakov2021}
\bibinfo{author}{{Lisakov}, M.~M.} \emph{et~al.}
\newblock \bibinfo{title}{{An Oversized Magnetic Sheath Wrapping around the
  Parsec-scale Jet in 3C 273}}.
\newblock \emph{\bibinfo{journal}{\apj}} \textbf{\bibinfo{volume}{910}},
  \bibinfo{pages}{35} (\bibinfo{year}{2021}).

\end{thebibliography}

\end{document}